\documentclass[preprint,12pt]{elsarticle}

\usepackage[utf8]{inputenc}
\usepackage[T1]{fontenc}
\usepackage{graphicx}
\usepackage{amsmath,amssymb,amsthm}
\usepackage{booktabs}
\usepackage{threeparttable}
\usepackage{longtable}
\usepackage{float}
\usepackage{array}
\usepackage{enumitem}
\usepackage{siunitx}
\DeclareSIUnit{\Molar}{M}
\usepackage[hidelinks]{hyperref}
\hypersetup{
  pdftitle={Auditing Haldane Consistency in Reversible Enzyme Kinetics: A Curated Two-Sided Backbone and a Labeled Fold-Error Benchmark},
  pdfauthor={Megan Simons and Jonathan Washburn},
  pdfsubject={Reversible enzyme kinetics; biochemical thermodynamics; Haldane consistency},
  pdfkeywords={Haldane relation; enzyme kinetics; biochemical thermodynamics; data curation; semi-synthetic benchmark}
}
\usepackage{rotating}
\usepackage{pdflscape}
\usepackage{placeins}
\usepackage{needspace}
\usepackage{xcolor}
\usepackage{microtype}

\setlist[itemize]{topsep=3pt,itemsep=2pt,parsep=0pt}
\setlist[enumerate]{topsep=3pt,itemsep=2pt,parsep=0pt}

\newtheorem{theorem}{Theorem}

\theoremstyle{definition}

\theoremstyle{remark}

\newcommand{\Keq}{K'_{\mathrm{eq}}}
\newcommand{\Kkin}{K'_{\mathrm{eq,kin}}}
\newcommand{\Kthermo}{K'_{\mathrm{eq,thermo}}}
\newcommand{\kcatf}{k_{\mathrm{cat}}^{+}}
\newcommand{\kcatr}{k_{\mathrm{cat}}^{-}}
\newcommand{\KMS}{K_{M,S}}
\newcommand{\KMP}{K_{M,P}}
\newcommand{\Jcost}{J}
\newcommand{\CH}{C_{\mathrm{Haldane}}}
\newcommand{\dG}{\Delta_r G^{\prime\circ}}
\newcommand{\Rpos}{\mathbb{R}_{>0}}

\makeatletter
\def\ps@pprintTitle{%
  \let\@oddhead\@empty
  \let\@evenhead\@empty
  \def\@oddfoot{}%
  \let\@evenfoot\@oddfoot}
\makeatother

\begin{document}

\begin{frontmatter}

\title{Auditing Haldane Consistency in Reversible Enzyme Kinetics: A Curated
Two-Sided Backbone and a Labeled Fold-Error Benchmark}

\author[inst1]{Megan Simons\corref{cor1}}
\ead{msimons@recognitionphysics.org}
\author[inst1]{Jonathan Washburn}
\cortext[cor1]{Corresponding author.}
\address[inst1]{Recognition Science, Recognition Physics Institute, Austin, TX, USA}

\begin{abstract}
Reversible enzyme kinetic constants can be audited through the Haldane relation: the
apparent equilibrium constant implied by the rate law should match biochemical thermodynamics under
matched conditions. We use the reciprocal cost $\CH=\Jcost(\Kkin/\Kthermo)$, with
$\Jcost(x)=\tfrac12(x+x^{-1})-1=\cosh(\ln x)-1$, as a calibrated, direction-symmetric reporting scale.
The score is zero at agreement, penalizes reciprocal over- and underestimates equally, encodes the
free-energy discrepancy in $RT$ units, and ranks records identically to $|\Delta\Delta G|$; the contribution
is therefore biochemical curation, a reproducible workflow, and fold-error calibration rather than a new
ordering. We apply the score to a curated demonstration set and, under prespecified inclusion criteria,
assemble a two-sided backbone of twenty-one audited single-study records. Eight genuinely independent tests
pair kinetics fit without a thermodynamic prior against separately measured equilibria; all eight fall within
twofold (maximum $\CH=0.069$), although this remains a feasibility demonstration. Across the full backbone,
eighteen records fall within twofold and three are flagged. The backbone is concentrated in carbohydrate
isomerases and epimerases, so these results are within-family observations. Because real records carry no
ground-truth labels, a semi-synthetic benchmark (twenty-nine within-twofold seeds, $1{,}885$ injected
known-error cases) quantifies detectability: AUC $0.784$ ($95\%$ bootstrap CI $0.725$--$0.838$),
conditional on the injected error taxonomy and invariant under monotone rescaling of $|\ln x|$. All data,
code, protocol, and benchmark generator are archived for exact reproduction.
\end{abstract}


\end{frontmatter}

% ============================================================
\section{Introduction}
\label{sec:intro}

\subsection{Thermodynamic consistency in reversible enzyme kinetics}

Reversible enzyme kinetic constants provide a quantitative meeting point between kinetics and
biochemical thermodynamics. For a reversible reaction operating in a closed system, microscopic
reversibility (detailed balance) ties the kinetic constants to the apparent equilibrium constant of
the catalyzed reaction. Auditing that link is a biophysical-chemistry problem: it asks whether rate
constants, Michaelis constants, reaction normalization, pH, temperature, ionic strength, and buffer
conditions describe the same transformed biochemical reaction. In the simplest one-substrate,
one-product mechanism,
\begin{equation}
E+S \;\rightleftharpoons\; ES \;\rightleftharpoons\; E+P,
\end{equation}
the four kinetic constants of the reversible Michaelis--Menten rate law are constrained by the Haldane
relation~\cite{Haldane1930,Cleland1963,CornishBowden2012}, which fixes the apparent equilibrium
constant implied by the kinetics,
\begin{equation}
\label{eq:haldane-intro}
\Kkin=\frac{\kcatf/\KMS}{\kcatr/\KMP}.
\end{equation}
The same reaction also has an apparent equilibrium constant determined by biochemical thermodynamics,
either measured directly or estimated from group- and component-contribution methods. Throughout,
$\Keq$ (read ``$K$-prime'') denotes this apparent (transformed) equilibrium constant evaluated at a
specified pH, temperature, ionic strength, and free magnesium concentration, in the IUBMB/NIST
convention of Alberty~\cite{Alberty2003} and the TECRDB compilations~\cite{Goldberg2004,GoldbergTewari1995Iso}.
If the kinetic constants and the thermodynamics describe the same reaction under the same conditions,
the two values of $\Keq$ must agree: they are two independent windows onto a single reaction-level
quantity. When they disagree, something in the data, the metadata, or the modeling assumptions
requires scrutiny.

To keep the Haldane expression unambiguous, the core analysis is restricted to reversible uni--uni
reactions $S\rightleftharpoons P$, or to pseudo-uni--uni reactions in which cosubstrate concentrations
are fixed and absorbed into effective constants. Multi-substrate mechanisms require mechanism-specific
Haldane relations~\cite{Cleland1963}; we treat several such reactions (ordered, rapid-equilibrium
random, and ping-pong bi--bi) explicitly, showing that the score is unchanged by this step.

\subsection{Heterogeneity of kinetic and thermodynamic records}

Enzyme data are abundant but heterogeneous. Kinetic constants for a given reaction may be reported for
one enzyme preparation, organism, isoform, mutant, pH, temperature, buffer, or assay direction, while
the equilibrium constant or reaction Gibbs energy may be reported for another. The relevant public
resources were built for different purposes: the Thermodynamics of Enzyme-Catalyzed Reactions Database
(TECRDB) collects apparent equilibrium constants and calorimetric enthalpies~\cite{Goldberg2004};
SABIO-RK~\cite{Wittig2012,Wittig2018} and BRENDA~\cite{Brenda2026} catalog kinetic parameters at
scale; and eQuilibrator supplies component-contribution estimates of transformed Gibbs
energies~\cite{Flamholz2012,Beber2022,Noor2013}. What is scarce is the pairing the Haldane
relation requires: forward and reverse kinetics for the same enzyme, reaction, and conditions, reported
in a single study, alongside a condition-matched equilibrium constant. To our knowledge, no public
corpus has been curated to that single-study two-sided standard. Rate-constant sets from
thermodynamically constrained global fits are a separate category: because they impose the equilibrium
constant during fitting, they satisfy the Haldane relation by construction and are not independent audits.

\subsection{Per-reaction Haldane discrepancy scoring and benchmarking}
\label{sec:contrib}

Thermodynamic consistency is routinely imposed in network-scale kinetic modeling --- through
parameter balancing~\cite{Liebermeister2006,Lubitz2010,Lubitz2019}, thermodynamically consistent
parameterization~\cite{SaaNielsen2015}, and reconciliation of inconsistent
rate-constant sets~\cite{Zielinski2024} --- and feasibility of a whole flux/concentration distribution
is tested by network-embedded thermodynamic analysis and thermodynamics-based metabolic flux
analysis~\cite{Kummel2006,Henry2007}. What is comparatively underdeveloped is a simple, calibrated,
per-reaction reporting scale for the single-record Haldane discrepancy itself, together with a
reproducible workflow and a benchmark of its behavior as a fold-band classifier. This paper
supplies all three, in four parts:
\begin{enumerate}
  \item We apply the canonical reciprocal cost of Washburn and Zlatanovi\'c~\cite{WashburnZlatanovic2026}
  as a direction-symmetric Haldane-consistency score $\CH=\Jcost(\Kkin/\Kthermo)$. The functional form is
  taken from that work; our contribution is the biochemical curation, benchmarking, and protocol. We give
  the score's calibration, restate the five-axiom uniqueness characterization, and use it as a
  calibrated reporting scale rather than a new record-ranking criterion (Section~\ref{sec:score}).
  \item We specify a reproducible curation protocol and apply the score to a curated
  demonstration set: two independent isomerase tests (phosphoglucose isomerase and triosephosphate
  isomerase), a buffer-sensitive effective-uni--uni fumarase case, seven racemase reciprocal-symmetry
  controls, and one worked ordered bi--bi example
  (Sections~\ref{sec:curation}--\ref{sec:poc}).
  \item Under inclusion criteria, an error taxonomy, and a minimum record threshold all fixed before
  the harvest, we assemble a real two-sided backbone of twenty-one audited single-study records
  --- sixteen uni--uni and five bi--bi spanning all three canonical bi--bi mechanisms, eighteen within
  twofold and three flagged, eight genuinely independent --- which supplies a descriptive empirical
  reference distribution for the curated sample (Section~\ref{sec:backbone}).
  \item Because real records carry no ground-truth error labels, we benchmark the fixed fold-band classifier
  on a semi-synthetic labeled dataset, on which the score attains an area under the receiver operating
  characteristic (ROC) curve of $0.784$
  and identifies its two least-detectable error modes; these operating characteristics are conditional on the
  injected error model rather than external misclassification rates (Section~\ref{sec:benchmark}).
\end{enumerate}
The Supplementary Material collects the full uniqueness proof, the record-by-record harvest narrative,
the long Standards for Reporting Enzyme Data (STRENDA) metadata, provenance, and per-record data tables, the bi--bi mechanism derivations, the
comparator-sensitivity analyses, and the code/data verification outputs, so that the main text stays
concise while the Supplementary Material provides the full audit trail.

% ============================================================
\section{Haldane-consistency score}
\label{sec:score}

\subsection{Haldane relation for reversible uni--uni reactions}
\label{sec:haldane}

We restrict to reversible one-substrate, one-product reactions $S\rightleftharpoons P$ catalyzed by an
enzyme that obeys the reversible Michaelis--Menten rate law~\cite{CornishBowden2012}. Writing
$V_{\max}^{+}=\kcatf[E]_{\mathrm{tot}}$ and $V_{\max}^{-}=\kcatr[E]_{\mathrm{tot}}$ (where
$[E]_{\mathrm{tot}}$ is the total enzyme concentration) for the forward and
reverse limiting rates and $\KMS,\KMP$ for the Michaelis constants of substrate and product, the net rate
can be written
\begin{equation}
\label{eq:rmm-rate}
v=\frac{(V_{\max}^{+}/\KMS)[S]-(V_{\max}^{-}/\KMP)[P]}
{1+[S]/\KMS+[P]/\KMP}.
\end{equation}
At equilibrium $v=0$, so the positive denominator drops out and the numerator gives
\[
\frac{V_{\max}^{+}}{\KMS}[S]_{\mathrm{eq}}
=\frac{V_{\max}^{-}}{\KMP}[P]_{\mathrm{eq}}.
\]
Thus the apparent equilibrium constant implied by the kinetics is
$[P]_{\mathrm{eq}}/[S]_{\mathrm{eq}}=(V_{\max}^{+}\KMP)/(V_{\max}^{-}\KMS)$, which in turnover form is
the Haldane relation
\begin{equation}
\label{eq:haldane-final}
\boxed{\;
\Kkin=\frac{\kcatf\,\KMP}{\kcatr\,\KMS}=\frac{\kcatf/\KMS}{\kcatr/\KMP}\; }.
\end{equation}
The right-hand form expresses $\Kkin$ as the ratio of forward and reverse specificity constants
$\kcatf/\KMS$ and $\kcatr/\KMP$~\cite{Cleland1963,CornishBowden2012}. We use $k_{\mathrm{cat}}$ rather
than $V_{\max}$ wherever possible; $V_{\max}$ values are admitted only when the enzyme concentration
and assay normalization are demonstrably comparable between the forward and reverse measurements, since
otherwise the cancellation leading to~\eqref{eq:haldane-final} fails. The appearance of $\KMS$ and
$\KMP$ follows from the reversible rate-law coefficients and does not assume that $K_M$ is a
thermodynamic dissociation constant. For pseudo-uni--uni reactions, a fixed cosubstrate or solvent
activity is absorbed into the effective rate-law constants and into the apparent transformed
equilibrium constant; thus a hydratase written as fumarate $+$ H$_2$O $\rightleftharpoons$ malate can be
audited as an effective fumarate $\rightleftharpoons$ malate transformation only under the biochemical
convention in which water activity is fixed. Multi-substrate records are scored through the
mechanism-specific Haldane relation~\cite{Cleland1963} (Section~\ref{sec:backbone};
derivations in the Supplementary Material), while the score below is unchanged.

\subsection{Thermodynamic equilibrium comparator}
\label{sec:thermo}

The thermodynamic comparator is the apparent equilibrium constant obtained from the transformed reaction
Gibbs energy under the stated biochemical standard-state convention and at the relevant temperature, pH,
ionic strength, and free magnesium
concentration~\cite{Alberty2003},
\begin{equation}
\label{eq:thermo}
\Kthermo=\exp\!\left(-\frac{\dG_{\mathrm{thermo}}}{RT}\right),
\qquad
\dG_{\mathrm{thermo}}=-RT\ln\Kthermo,
\end{equation}
with $R$ the gas constant and $T$ the absolute temperature. Analogously $\dG_{\mathrm{kin}}\equiv
-RT\ln\Kkin$, so the free-energy discrepancy is $\Delta\Delta G=\dG_{\mathrm{kin}}-\dG_{\mathrm{thermo}}$.
We use a primary experimental comparator from TECRDB matched as closely as possible in $T$, pH, ionic
strength, and free magnesium activity (pMg)~\cite{Goldberg2004}, and a secondary eQuilibrator estimate adjusted to the reported
conditions~\cite{Beber2022,Noor2013}.

\paragraph{Racemase controls as symmetry references}
For free enantiomers in an achiral medium under identical standard-state conventions (for example
$(S)$-mandelate $\rightleftharpoons (R)$-mandelate, or L-proline $\rightleftharpoons$ D-proline),
substrate and product are mirror images with identical standard Gibbs energies of formation, so
$\dG=0$ and $\Kthermo=1$ by symmetry; the enzyme (including any pyridoxal 5$'$-phosphate (PLP) cofactor) is a catalyst and does
not alter the free-reaction equilibrium constant. Such reactions furnish a parameter-free reference: the
score below collapses to $\CH=\Jcost(\Kkin)$, so any departure of $\Kkin$ from unity reflects the
reciprocity of the reported directional kinetic terms. Seven racemase records serve as internal
controls under this symmetry convention (Section~\ref{sec:poc}).

\subsection{Reciprocal Haldane-consistency cost}
\label{sec:cost}

\begin{figure}[!htbp]
\centering
\includegraphics[width=0.78\textwidth]{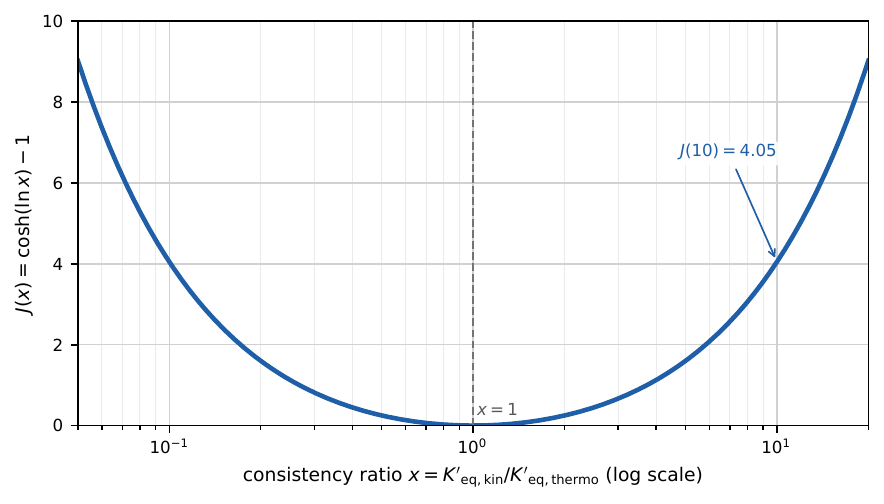}
\caption{Shape of the reciprocal cost $\Jcost(x)=\tfrac12(x+x^{-1})-1=\cosh(\ln x)-1$ on a log-scaled
ratio axis. The score is zero at exact agreement ($x=1$), symmetric under reciprocal error,
near-quadratic in $\ln x$ near the minimum, and grows steeply for large multiplicative disagreement; the
twofold reporting boundary used below is $\Jcost(2)=0.25$.}
\label{fig:Jshape}
\end{figure}

Let the consistency ratio be $x=\Kkin/\Kthermo\in\Rpos$, and define the score as the reciprocal cost
\begin{equation}
\label{eq:CHdef}
\CH=\Jcost(x)=\tfrac12\bigl(x+x^{-1}\bigr)-1=\frac{(x-1)^2}{2x},
\end{equation}
the last (``perfect-square'') form making non-negativity explicit: $\Jcost(x)\ge0$ for $x>0$, with
equality only at $x=1$. We refer to $\CH$ as a score in the audit context and reserve the term cost for
$\Jcost$ when discussing its axiomatic properties. The elementary properties that make $\CH$ a useful
consistency score are:
\begin{itemize}
  \item \emph{Exact agreement.} $\CH=0\Leftrightarrow x=1\Leftrightarrow\Kkin=\Kthermo$.
  \item \emph{Reciprocal symmetry.} $\Jcost(x^{-1})=\Jcost(x)$, so an $n$-fold overestimate and an
  $n$-fold underestimate receive identical cost; the score privileges neither the kinetic nor the thermodynamic side, nor the
  reaction direction.
  \item \emph{Free-energy form.} With $\delta=\ln x$ we have $\cosh\delta=\tfrac12(x+x^{-1})$, and since
  $\dG_{\mathrm{kin}}=-RT\ln\Kkin$ and $\dG_{\mathrm{thermo}}=-RT\ln\Kthermo$, the free-energy discrepancy
  is $\Delta\Delta G=\dG_{\mathrm{kin}}-\dG_{\mathrm{thermo}}=-RT\delta$. Because $\cosh$ is even,
  \begin{equation}
  \label{eq:CHfree}
  \boxed{\;\CH=\cosh(\delta)-1=\cosh\!\left(\frac{\Delta\Delta G}{RT}\right)-1\;},
  \end{equation}
  where $\Delta\Delta G$ is defined here as $\dG_{\mathrm{kin}}-\dG_{\mathrm{thermo}}=-RT\delta$. Thus the
  score is a hyperbolic cosine of the transformed free-energy discrepancy in units of $RT$, and the
  signed discrepancy is reported separately as $\delta=-\Delta\Delta G/RT$. At
  \SI{298.15}{\kelvin} ($RT\approx\SI{2.479}{\kilo\joule\per\mole}$), a twofold ratio corresponds to
  $|\Delta\Delta G|\approx\SI{1.72}{\kilo\joule\per\mole}$ and a tenfold ratio to
  $|\Delta\Delta G|\approx\SI{5.71}{\kilo\joule\per\mole}$.
  \item \emph{Inverse calibration.} The calibration inverts in closed form, so a reported score reads
  directly as a fold error and a free-energy gap: $|\ln x|=\operatorname{arcosh}(1+\CH)$,
  $F\equiv\max(x,x^{-1})=\exp[\operatorname{arcosh}(1+\CH)]$, and
  $|\Delta\Delta G|=RT\,\operatorname{arcosh}(1+\CH)$. The sign of the discrepancy is reported
  separately as $x$ or $\delta=\ln x$.
  \item \emph{Small-error behavior and one-sided bound.} The Taylor series
  $\cosh\delta-1=\tfrac12\delta^2+\tfrac{1}{24}\delta^4+O(\delta^6)$ gives
  $\CH=\tfrac12\delta^2+\tfrac{1}{24}\delta^4+O(\delta^6)$,
  so near agreement $\CH\approx\tfrac12(\ln x)^2$ recovers the familiar squared-log-error penalty;
  because every higher term is non-negative, $\CH\ge\tfrac12(\ln x)^2$ globally, and the score grows
  like $\tfrac12 e^{|\delta|}$ for large multiplicative disagreement.
\end{itemize}
Table~\ref{tab:fold} lists representative fold errors, and Figure~\ref{fig:Jshape} plots $\Jcost(x)$:
a unique minimum of zero at $x=1$, exact symmetry under $x\leftrightarrow1/x$, near-quadratic behavior
in $\ln x$, and a steep $\cosh$ penalty for large fold errors.

\begin{table}[!htbp]
\centering
\caption{Fold-error interpretation of the reciprocal cost $\Jcost$. Logarithms are natural; the
$|\ln x|$ column is to base $e$. By reciprocal symmetry each entry applies equally to a factor of $x$
and to its reciprocal $1/x$.}
\label{tab:fold}
\begin{tabular}{ccc}
\toprule
Fold error $x$ or $1/x$ & $|\ln x|$ & $\Jcost(x)$ \\
\midrule
$1$   & $0.000$ & $0.000$ \\
$2$   & $0.693$ & $0.250$ \\
$3$   & $1.099$ & $0.667$ \\
$4$   & $1.386$ & $1.125$ \\
$5$   & $1.609$ & $1.600$ \\
$10$  & $2.303$ & $4.050$ \\
$100$ & $4.605$ & $49.005$ \\
\bottomrule
\end{tabular}
\end{table}

\subsection{Fold-discrepancy reporting bands}
\label{sec:bands}

For descriptive classification, the continuous score is grouped into round fold-error bands
(Table~\ref{tab:bands}). These are fixed fold-discrepancy reporting conventions, neither tuned to any
dataset nor validated as operating points of a decision procedure; the continuous $\CH$ remains the
primary reported quantity, and the band cut points are the fixed free-energy gaps $RT\ln2$, $RT\ln5$,
and $RT\ln10$, invariant under any monotone rescaling of $|\ln x|$. Estimating how this banding performs
as a classifier is the aim of Section~\ref{sec:benchmark}.

\begin{table}[!htbp]
\centering
\caption{Descriptive raw-score bands for the Haldane-consistency score. The bands are fixed
fold-discrepancy ranges for reporting $\CH$, not validated decision categories. At \SI{298.15}{\kelvin}
the boundaries correspond to $|\Delta\Delta G|=RT\ln 2$, $RT\ln 5$, and $RT\ln 10$ ($\approx1.72$,
$3.99$, and \SI{5.71}{\kilo\joule\per\mole}); the upper numeric boundary is
$4.05=\cosh(\ln10)-1$.}
\label{tab:bands}
\begin{tabular}{@{}>{\raggedright\arraybackslash}p{0.26\textwidth}>{\raggedright\arraybackslash}p{0.34\textwidth}>{\centering\arraybackslash}p{0.20\textwidth}@{}}
\toprule
Fold-discrepancy band & Ratio condition & $\CH$ range \\
\midrule
within twofold & $1/2 \le x \le 2$ & $\le 0.25$ \\
2--5-fold discrepancy & $2 < x \le 5$ or $1/5 \le x < 1/2$ & $0.25<\CH\le1.60$ \\
5--10-fold discrepancy & $5 < x \le 10$ or $1/10 \le x < 1/5$ & $1.60<\CH\le4.05$ \\
$>$tenfold discrepancy & $x>10$ or $x<1/10$ & $>4.05$ (unbounded above) \\
\bottomrule
\end{tabular}
\end{table}

\subsection{Axiomatic characterization and calibration}
\label{sec:uniqueness}

The reciprocal score is used here as a biochemical audit scale, but its form is also characterized by a
short uniqueness statement, established by Washburn and Zlatanovi\'c~\cite{WashburnZlatanovic2026} and
restated here for the positive equilibrium-constant ratio. A function $C:\Rpos\to\mathbb{R}$ is a
\emph{calibrated reciprocal cost} if it satisfies: (C1) reciprocal symmetry, $C(x)=C(x^{-1})$; (C2)
normalization, $C(1)=0$; (C3) the composition law $C(uv)+C(u/v)=2C(u)C(v)+2C(u)+2C(v)$ for all
$u,v>0$; (C4) continuity; and (C5) unit calibration, $c''(0)=1$ for $c(t)=C(e^{t})$.

\begin{theorem}[Uniqueness of the calibrated reciprocal cost~\cite{WashburnZlatanovic2026}]
\label{thm:uniqueness}
A function $C:\Rpos\to\mathbb{R}$ satisfies axioms \textup{(C1)--(C5)} if and only if
$C(x)=\Jcost(x)=\tfrac12(x+x^{-1})-1$.
\end{theorem}

Writing $H=C+1$ turns axiom~(C3) into the multiplicative d'Alembert equation
$H(uv)+H(u/v)=2H(u)H(v)$~\cite{Aczel1966}, whose
continuous solutions in the logarithmic coordinate are $\cosh(\alpha t)$, $\cos(\beta t)$, or the
constant; the positive sign of the calibration selects the hyperbolic branch, and the unit magnitude
$c''(0)=1$ fixes $\alpha=1$. Without that normalization convention, the same d'Alembert family would
contain $\cosh(\alpha\ln x)-1$ with other positive curvatures. The three algebraic axioms encode comparison rather than enzyme mechanism: (C1) makes the
cost direction-symmetric, (C2) makes exact agreement cost-free, and (C3) imposes a reciprocal
composition rule that distinguishes $\Jcost$ from the naive half-squared log-error
$\tfrac{1}{2}(\ln x)^{2}$, which shares the same unit calibration $c''(0)=1$ but does not
satisfy (C3). The complete d'Alembert classification and proof are
given in the Supplementary Material.

The axioms \textup{(C1)--(C5)} and Theorem~\ref{thm:uniqueness} are imported unchanged from Washburn and
Zlatanovi\'c~\cite{WashburnZlatanovic2026}: in the present biochemical application (C3) is a mathematical
selection principle that picks $\Jcost$ out of the d'Alembert family and (C5) is a scale normalization,
neither an enzyme-mechanistic law nor a claim of biochemical privilege for $\CH$. The biochemical claims
below therefore rely only on the score's symmetry, fold-error calibration, and monotonicity. Because
$\CH$ is strictly increasing in $|\ln x|$, it preserves the ordering of records by $|\ln x|$ and
$|\Delta\Delta G|$ and is a calibrated reporting scale rather than a new record-ranking criterion. Its
practical distinction from the equally symmetric $|\ln x|$, $(\ln x)^2$, and $|\Delta\Delta G|$ is the
free-energy calibration with closed-form inverse, the one-sided bound $\CH\ge\tfrac12(\ln x)^2$, and the
composition identity (C3) (Table~\ref{tab:scores}).

\begin{table}[!htbp]
\centering
\caption{Symmetry and scale of alternative kinetic--thermodynamic discrepancy scores. Here
$|\Delta\Delta G|=RT\,|\ln x|$ and $F=\max(x,x^{-1})$ is the reciprocal-symmetric fold discrepancy.
The quantities $|\ln x|$, $(\ln x)^2$, and $F-1$ are dimensionless like $\CH$: $\CH$ is neither uniquely
symmetric nor uniquely dimensionless. Its local distinction from squared log error is the higher-order
tail: $\CH=\tfrac12(\ln x)^2+O((\ln x)^4)$ near agreement.}
\label{tab:scores}
\scriptsize
\setlength{\tabcolsep}{4pt}
\begin{tabular}{@{}>{\raggedright\arraybackslash}p{4.2cm}
                >{\centering\arraybackslash}p{3.4cm}
                >{\centering\arraybackslash}p{1.9cm}
                >{\raggedright\arraybackslash}p{3.3cm}@{}}
\toprule
Score & Formula & Symmetric under $x\to 1/x$? & Reported scale \\
\midrule
Absolute fold error & $|x-1|$ & No$^{a}$ & dimensionless ratio \\
Symmetric fold discrepancy & $F-1,\ F=\max(x,x^{-1})$ & Yes & dimensionless fold \\
Log error & $|\ln x|$ & Yes & dimensionless log-ratio \\
Squared log error & $(\ln x)^2$ & Yes & dimensionless log-ratio \\
Free-energy error & $|\Delta\Delta G|$ & Yes & energy ($RT$ units) \\
Reciprocal Haldane-consistency score & $\cosh(\ln x)-1$ & Yes & dimensionless cost \\
Small-error limit of $\CH$ & $\tfrac12(\ln x)^2+O((\ln x)^4)$ & Yes & local squared-log scale \\
\bottomrule
\end{tabular}
\begin{minipage}{\linewidth}
\vspace{3pt}\footnotesize
$^{a}$~$|x-1|$ penalizes an $n$-fold overestimate ($x=n$) more strongly than an $n$-fold underestimate
($x=1/n$); all other entries are symmetric under reciprocal exchange. The symmetric fold discrepancy
$F-1$ is the directly interpretable reciprocal-symmetric fold measure; $\CH$ additionally provides
the free-energy calibration and the structural properties noted in the text.
\end{minipage}
\end{table}

\paragraph{Uncertainty-aware reporting}
Kinetic and thermodynamic estimates carry measurement uncertainty. Working in the log-ratio coordinate
$\delta=\ln x$ and treating the kinetic and thermodynamic log-estimates as independent gives
$\sigma_\delta^{2}=\sigma^{2}_{\ln\Kkin}+\sigma^{2}_{\ln\Kthermo}$, with the kinetic term following the
additive decomposition $\ln\Kkin=\ln\kcatf+\ln\KMP-\ln\kcatr-\ln\KMS$. Where a meaningful spread is
available we report the standardized log-deviation $z=\delta/\sigma_\delta$ alongside the continuous
$\CH$; where a joint fit covariance is unavailable we bracket its effect rather than assume
independence. The full propagation, including the correlation bracket used for the bi--bi records, is
given in the Supplementary Material.

% ============================================================
\section{Curation protocol and inclusion criteria}
\label{sec:curation}

\subsection{Study design}

\begin{figure}[!htbp]
\centering
\includegraphics[width=0.82\textwidth]{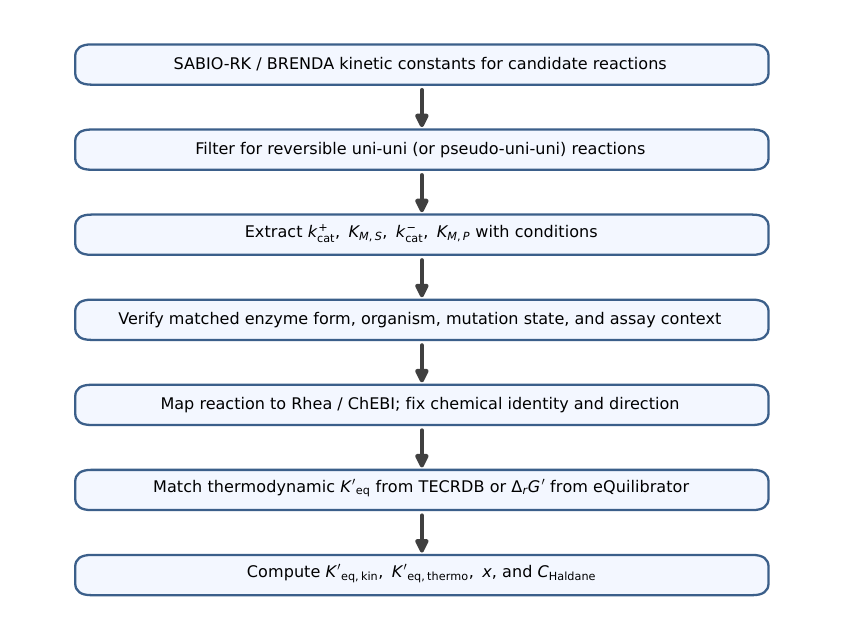}
\caption{Data curation and scoring workflow for reversible uni--uni reactions. Kinetic and
thermodynamic records are independently retrieved, normalized to a common reaction identity, checked
for matched enzyme form and assay context, and combined into the reciprocal Haldane-consistency score;
arrows indicate the curation order. Both record sets follow this structure, with manual primary-source
extraction used when database records alone are not sufficiently matched.}
\label{fig:workflow}
\end{figure}

The analysis combines two record sets curated under the same auditable workflow
(Figure~\ref{fig:workflow}). The first is a small manually curated demonstration set
(Section~\ref{sec:poc}), assembled to cover selected individual records rather than by a systematic
search; it demonstrates the score's behavior on those records without providing a standalone score distribution or
misclassification rate. The second is a prespecified real two-sided backbone
(Section~\ref{sec:backbone}), harvested under inclusion criteria, an error taxonomy, and a minimum
record threshold fixed before the harvest, and paired with a semi-synthetic labeled benchmark
(Section~\ref{sec:benchmark}). For each record, a single internally consistent source supplies the
constants needed for the Haldane calculation together with the available experimental metadata; missing
metadata are recorded explicitly and treated as a limitation rather than inferred.

The workflow in Figure~\ref{fig:workflow} defines the practical unit of analysis used throughout the
paper: a kinetic record enters the score only after reaction identity, assay context, and comparator
conditions have been aligned. This alignment step is what separates the curated demonstration records
from the larger backbone harvest, where many otherwise relevant enzyme records are excluded because one
side of the two-sided comparison is missing or not condition matched.

\subsection{Data sources and reaction normalization}

For scalable use, kinetic constants are drawn preferentially from SABIO-RK, which stores rate laws,
parameters, and conditions in a reaction-centered curated form~\cite{Wittig2012,Wittig2018}, with
BRENDA as a broad enzyme-centered complement~\cite{Brenda2026}; where neither keeps the forward and
reverse constants within a single enzyme form and assay context, we extract manually from the primary
study. Thermodynamic comparators are taken from TECRDB as the primary experimental
source~\cite{Goldberg2004} and from eQuilibrator's component-contribution framework as a secondary
estimate and for condition adjustments~\cite{Beber2022,Noor2013}. Reactions are mapped to Rhea
identifiers, built on the ChEBI chemical ontology~\cite{Bansal2022,Hastings2016}, so that the kinetic
and thermodynamic records refer to the same chemical transformation, protonation and hydration
convention, cofactor set, and direction before any comparison; kinetic metadata follow the STRENDA
reporting standards~\cite{Tipton2014} and the records are shared in a Findable, Accessible,
Interoperable, and Reusable (FAIR) form~\cite{Wilkinson2016}.

\subsection{Inclusion and exclusion criteria}
\label{sec:inclusion}

A two-sided kinetic-versus-thermodynamic test is included only if: the reaction is reversible and can be
treated as uni--uni (true, or pseudo-uni--uni with fixed cosubstrates absorbed into effective constants),
or a specified mechanism-specific multi-substrate Haldane relation applies;
all reported kinetic constants correspond to the same enzyme form, organism, mutation state, and assay
conditions; forward and reverse directions are unambiguously defined and consistently oriented; the
kinetic constants are positive and convertible to common units; sufficient condition metadata are
available to justify the comparison; the thermodynamic reaction matches the kinetic reaction after
Rhea/ChEBI normalization and is condition-matched or explicitly condition-adjusted; and no allostery,
cooperativity, substrate inhibition, multi-substrate ambiguity, or irreversible-assay assumption
contaminates the Haldane calculation. Records are excluded for missing reverse constants, ambiguous
substrate/product identity, mixed isoforms or mutants across the two directions, mismatched pH or
temperature without a defensible correction, incompatible mechanism, unreconstructable
saturating-cosubstrate assumptions, or a thermodynamic comparator that cannot be aligned to the kinetic
reaction. Thermodynamically constrained global fits are also excluded from the backbone because the
fitted constants inherit the imposed equilibrium constraint and therefore carry no independent Haldane
test. The seven racemase controls instead use the symmetry reference $\Kthermo=1$
(Section~\ref{sec:thermo}) and serve only as reciprocal-symmetry checks on the reported kinetics.

\subsection{Computation and reporting}

The assembled kinetic equilibrium constant is computed in log space,
$\ln\Kkin=\ln\kcatf+\ln\KMP-\ln\kcatr-\ln\KMS$, from whichever of three algebraically equivalent input
forms a source reports: the full four-constant Haldane expression; matched forward/reverse specificity
constants $\kcatf/\KMS$ and $\kcatr/\KMP$ (or directional $V_{\max}/K_M$ terms, where a shared
active-enzyme normalization is demonstrated); or a directly reported assembled constant from a
reversible fit. The thermodynamic comparator is $\ln\Kthermo=\ln K'_{\mathrm{TECRDB}}$ or
$-\dG_{\mathrm{eQ}}/RT$. We report the continuous score $\CH$ as the primary quantity for every record;
the fold bands of Table~\ref{tab:bands} are descriptive categories, are not applied to uncertainty
summaries such as $z$, and were set by no receiver-operating-characteristic calibration.

In the backbone, a record is called a flagged inconsistency only when its central fold discrepancy exceeds
the twofold boundary and the provenance supports a record-level Haldane audit. Independence status is
reported separately: the eight independent tests pair kinetic constants fit without a thermodynamic prior
against separately measured equilibria, whereas same-study comparator records remain descriptive
condition-matched checks. For records assembled from multi-constant literature compilations with
unavailable joint covariance, the central $\CH$ is reported but the record is not promoted to evidence of
inconsistency unless the covariance bracket and provenance support that interpretation. We apply this as
an explicit bracket rule. A multi-constant bi--bi compilation such as the illustrative horse-liver ADH
assembles $\Kkin$ from six constants whose unknown but plausibly positive correlations shrink the
standardized log-deviation $z=\delta/\sigma_\delta$, so its $\approx2.35$-fold central deviation brackets
to a sub-$3\sigma$, unresolved discrepancy and is not flagged. The three flagged records, by contrast, are
uni--uni epimerase/isomerase audits whose flag does not rest on a covariance-dependent $z$: their central
fold discrepancies are fixed by the reported constants and remain above twofold under any correlation
assumption ($2.88$ for the \emph{C.\ scindens} DPEase, $3.09$ for pea RPI, and $3.96$ for the
\emph{Sinorhizobium} DTEase). For the \emph{C.\ scindens} psicose epimerase the discrepancy is localized
to the reported kinetics rather than the comparator, because congeners catalyzing the identical
D-fructose$\rightarrow$D-psicose epimerization are themselves within twofold (Supplementary Table~S7).
The \emph{Sinorhizobium} tagatose epimerase has no same-reaction congener in the backbone; it instead
retains its flag against an independently measured tagatose/sorbose equilibrium that stays above twofold
across the plausible comparator range. This is why the illustrative horse-liver
ADH compilation (fold $\approx2.35$) is not flagged, whereas the same-study \emph{C.\ scindens} DPEase
record (fold $=2.88$) is flagged: the latter is an exactly uni--uni central-value audit with a same-study
equilibrium comparator and a consistent congener on the same reaction, with no multi-constant covariance
freedom that could bring it within twofold. Where the per-constant joint covariance needed for a
standardized $z$ is unavailable, such a record is reported as a central-value curation target rather than a
statistically resolved inconsistency.

\subsection{Prespecified benchmark design}
\label{sec:design}

Because eligible two-sided records are scarce, the backbone is harvested under a protocol fixed in advance
as an internal analysis plan, which guards against retrofitting the criteria to the data. This
prespecification is internal: the analysis plan was frozen in the project repository before the harvest
and is archived with the release (Data and Code Availability), but it was not publicly
preregistered, and we use prespecified throughout in this internal sense.

We distinguish two layers of the protocol. The scientific core --- the inclusion and exclusion criteria above, the six-mode error
taxonomy of Section~\ref{sec:benchmark}, the scoring plan, and the two-, five-, and tenfold cuts ---
was fixed before the backbone harvest began and never altered. The candidate tracker,
the running list of enzymes to attempt, was by contrast expanded under those same fixed criteria as
harvesting revealed that single-study bidirectionality, not the score, was the limiting constraint: when
the first rounds showed eligible two-sided records to be scarce, later rounds widened the enzyme list (for
example to the rare-sugar ketose epimerases) without relaxing any acceptance rule. This protocol
amendment changed the search scope but not the analysis, because the criteria that decide consistency
were fixed throughout. No record was dropped, reclassified, or retained because of its computed
$\CH$ after the tracker was widened.

The protocol fixes an inclusion threshold: if fewer than about fifteen eligible two-sided records can be
assembled, the semi-synthetic benchmark is the primary instrument on the smaller real seed; once the
threshold is reached, the real backbone and the benchmark are analyzed together, the backbone
supplying a threshold-cleared empirical reference distribution for the score, with the labeled operating
characteristics still estimated against the injected taxonomy because the real records carry no
ground-truth error labels. After provisional inclusion, each record underwent a second independent audit:
parameter re-extraction from the primary source where available, re-computation of $\Kkin$, orientation
and unit checks, comparator cross-checks, and reaction-identifier verification. The scoring and benchmark
scripts are deterministic and standard-library Python, and the SHA-256 checksums of their regenerated
outputs are recorded with the archived release (Supplementary Section~S6), so the reported numerical
outputs are exactly reproducible.

% ============================================================
\section{Curated demonstration records}
\label{sec:poc}

The score is first applied to a small, carefully curated demonstration set of reversible
enzyme/reaction records for which the uni--uni Haldane relation~\eqref{eq:haldane-final} is appropriate
(Table~\ref{tab:curated}). Two records --- phosphoglucose isomerase (PGI) and triosephosphate isomerase
(TPI) --- are independent isomerase tests with measured thermodynamic comparators. Fumarase is retained
as a buffer-sensitive effective-uni--uni case: it has an independently measured comparator, but its main
signal is the phosphate-versus-non-phosphate contrast within one kinetic study. The remaining seven are
racemase controls for which $\Kthermo=1$ is fixed
by enantiomer symmetry. Reaction-normalization identifiers and full STRENDA-aligned kinetic metadata are
given in the Supplementary Material (Tables~S1--S3). The scored comparisons are plotted in
Figures~\ref{fig:poc-consistency}--\ref{fig:poc-distribution}.

\begin{table}[!htbp]
\centering
\caption{Curated enzyme/reaction records for the demonstration set. All records are
reversible and either genuinely uni--uni (isomerases, racemases) or pseudo-uni--uni with a
conventionally fixed cosubstrate (fumarase, water held constant). EC numbers and normalized reactions
follow the Rhea/ChEBI convention~\cite{Bansal2022,Hastings2016} (identifiers in Supplementary
Table~S1). Numerical scores are reported in Tables~\ref{tab:pgi},~\ref{tab:fumarase-kin},
and~\ref{tab:scored}.}
\label{tab:curated}
\scriptsize
\setlength{\tabcolsep}{3pt}
\begin{tabular}{@{}>{\raggedright\arraybackslash}p{2.8cm}>{\raggedright\arraybackslash}p{3.7cm}>{\raggedright\arraybackslash}p{3.1cm}>{\raggedright\arraybackslash}p{3.0cm}@{}}
\toprule
Enzyme (EC) & Reaction (Rhea-normalized) & Role in the analysis & Thermodynamic comparator \\
\midrule
Phosphoglucose isomerase (EC 5.3.1.9) &
glucose-6-phosphate $\rightleftharpoons$ fructose-6-phosphate &
Two-sided worked example &
Experimental $\Keq$ (TECRDB 60KAH/LOW) \\[2pt]
Fumarase (EC 4.2.1.2) &
fumarate $+$ H$_2$O $\rightleftharpoons (S)$-malate &
Scored kinetic Haldane case (buffer-sensitive) &
Experimental $\Keq\approx4.2$ (TECRDB) \\[2pt]
Triosephosphate isomerase (EC 5.3.1.1) &
glyceraldehyde-3-phosphate $\rightleftharpoons$ dihydroxyacetone phosphate &
Two-sided isomerase comparison &
Experimental $\Keq=22.0$~\cite{Veech1969} \\[2pt]
Mandelate racemase (EC 5.1.2.2) &
$(S)$-mandelate $\rightleftharpoons (R)$-mandelate &
$\Kthermo=1$ control (reciprocal symmetry) &
Symmetry reference ($\dG=0$) \\[2pt]
Proline racemase (EC 5.1.1.4) &
L-proline $\rightleftharpoons$ D-proline &
$\Kthermo=1$ control (second racemase) &
Symmetry reference ($\dG=0$) \\[2pt]
Glutamate racemase RacE2 (EC 5.1.1.3) &
L-glutamate $\rightleftharpoons$ D-glutamate &
$\Kthermo=1$ control (same-source bidirectional kinetics) &
Symmetry reference ($\dG=0$) \\[2pt]
Glutamate racemase RacE1 (EC 5.1.1.3) &
L-glutamate $\rightleftharpoons$ D-glutamate &
$\Kthermo=1$ control (isozyme contrast) &
Symmetry reference ($\dG=0$) \\[2pt]
Glutamate racemase (EC 5.1.1.3) &
L-glutamate $\rightleftharpoons$ D-glutamate &
$\Kthermo=1$ control (provenance-flagged kinetic record) &
Symmetry reference; reproduced, not re-extracted \\[2pt]
Aspartate racemase (EC 5.1.1.13) &
L-aspartate $\rightleftharpoons$ D-aspartate &
$\Kthermo=1$ control (reported $\Keq$ check) &
Symmetry reference ($\dG=0$) \\[2pt]
Alanine racemase DadX (EC 5.1.1.1) &
L-alanine $\rightleftharpoons$ D-alanine &
$\Kthermo=1$ control (PLP-dependent; provenance flagged) &
Symmetry reference; reproduced $V_{\max}/K_M$ \\
\bottomrule
\end{tabular}
\end{table}

Table~\ref{tab:curated} separates the roles of the demonstration records before any scoring is applied:
the two isomerases test condition-matched kinetic-versus-thermodynamic agreement, fumarase tests a
buffer-dependent effective-uni--uni contrast, and the racemases test reciprocal symmetry at
$\Kthermo=1$. The following subsections use those roles to interpret the individual scores rather than
treating the records as a single homogeneous sample.

\subsection{Phosphoglucose isomerase benchmark record}

Phosphoglucose isomerase provides a reference case because the same reaction has both a kinetically determined
equilibrium constant and source-traced evaluated thermodynamic constants. St\"odeman and Schwarz
measured the glucose-6-phosphate $\rightleftharpoons$ fructose-6-phosphate reaction on baker's yeast PGI
by differential stopped-flow microcalorimetry, reporting $\Kkin=0.307\pm0.053$ at
\SI{293.4}{\kelvin}~\cite{Stodeman2004}; the TECRDB-evaluated comparators (tracing to Kahana et
al.~\cite{Kahana1960} and Dyson and Noltmann~\cite{Dyson1968}) give $\Kthermo=0.260$--$0.290$. The
kinetic and thermodynamic estimates agree to within $1.2$--$1.4$-fold, so $\CH\le0.05$ and all pairings
are within twofold (Table~\ref{tab:pgi}). For the temperature- and pH-aligned pair (60KAH/LOW),
$x=1.18$ and $\CH=0.014$. With the single reported kinetic error bar ($\sigma_{\ln\Kkin}\approx0.17$,
obtained from $0.053/0.307\approx0.17$) and a two-comparator thermodynamic spread, the standardized discrepancy is $z\approx0.9$ on this
approximate resolution scale, reported as a descriptive sensitivity summary rather than a calibrated
test. eQuilibrator's component-contribution value for the same reaction
($K'_{\mathrm{eQ}}=0.342$)~\cite{Beber2022,Noor2013} scores $\CH^{\mathrm{est}}=0.006$ against the same
kinetic estimate, returning the same within-twofold band.

\begin{table}[!htbp]
\centering
\caption{Phosphoglucose isomerase, glucose-6-phosphate $\rightleftharpoons$ fructose-6-phosphate,
$\Keq=[\mathrm{F6P}]/[\mathrm{G6P}]$. Kinetic constants $\Kkin=0.307\pm0.053$ (\SI{293.4}{\kelvin}) and
$0.395\pm0.033$ (\SI{298.4}{\kelvin}) are from baker's-yeast PGI~\cite{Stodeman2004}; thermodynamic
comparators are evaluated TECRDB records~\cite{GoldbergTewari1995Iso} (codes tracing to
Kahana et al.~\cite{Kahana1960} and Dyson and Noltmann~\cite{Dyson1968}), temperature-matched to the
kinetic value. Here $x=\Kkin/\Kthermo$ and $\CH=\Jcost(x)$.}
\label{tab:pgi}
\small
\begin{tabular}{@{}>{\raggedright\arraybackslash}p{4.0cm}cccc>{\raggedright\arraybackslash}p{1.9cm}@{}}
\toprule
Thermodynamic comparator (TECRDB) & $\Kthermo$ & $\Kkin$ & $x$ & $\CH$ & Fold band \\
\midrule
60KAH/LOW, $293.15$\,K, pH $8.0$ & 0.260 & 0.307 & 1.18 & 0.014 & within twofold \\
68DYS/NOL, $298.2$\,K, pH $8.5$ & 0.290 & 0.395 & 1.36 & 0.048 & within twofold \\
\bottomrule
\end{tabular}
\end{table}

\subsection{Fumarase: a buffer-sensitive kinetic Haldane score}
\label{sec:fumarase}

For fumarase (EC 4.2.1.2), written as fumarate $+$ H$_2$O $\rightleftharpoons$ ($S$)-malate, the apparent
transformed comparator is $K'_{\mathrm{eq}}=4.20\pm0.05$ at \SI{298.15}{\kelvin} and near-neutral
pH~\cite{Gajewski1985,GoldbergTewari1995Lyases}; the representative value traces to TECRDB records
53BOC/ALB and 85GAJ/GOL, with the four independent TECRDB records spanning only $3.98$--$4.43$
(Supplementary Table~S4), so the comparator scatter is negligible at the resolution of the score. The
complete forward and reverse constants come from a single modern study: Genda et al.\ measured
\emph{Corynebacterium glutamicum} fumarase across six pH/buffer conditions~\cite{Genda2006}
(Supplementary Table~S5). Scored against $\Kthermo\approx4.2$ (Table~\ref{tab:fumarase-kin}), the
$100$~\si{\milli\Molar} phosphate-buffer rows are within twofold ($x=1.1$--$1.5$, $\CH\le0.09$) while the
non-phosphate (MES/TES/Tris) rows run $3.8$--$4.5$-fold high ($\CH=1.0$--$1.4$, the 2--5-fold band). This
result has two components. The within-pH phosphate-versus-non-phosphate contrast is
the primary, approximation-free signal: at each of pH~6, 7, and 8 the phosphate estimate is within
twofold while the non-phosphate estimate is several-fold high, a difference that sits directly in the raw
$k_{\mathrm{cat}}/K_M$ values. The absolute fold offset against $\Kthermo$, by contrast, admits two
distinct readings that the symmetric score flags but cannot separate: (i)~a genuine buffer-specific
physicochemical effect --- consistent with the buffer-dependent inhibition and $K_M$ shifts that Genda
et al.\ report for ATP and substrate analogs, and which they attribute mainly to effects on $K_M$
values --- which the score correctly registers as a real disagreement under those conditions; and
(ii)~the modeling assumption inherent in treating fumarase's multi-state catalytic cycle as effective
uni--uni, an effective-model limitation that no reciprocal Haldane comparison can resolve. Because the
reported constants do not distinguish the two, the non-phosphate flag is best read as a curation target
rather than a mechanism-level inconsistency.

\begin{table}[!htbp]
\centering
\caption{Kinetic Haldane scores for fumarase from the complete forward/reverse constants of
\emph{C. glutamicum} fumarase~\cite{Genda2006} (Supplementary Table~S5),
$\Kkin=(\kcatf\KMP)/(\kcatr\KMS)$, scored against the representative thermodynamic value
$\Kthermo=4.2$. Phosphate-buffer estimates are within twofold; non-phosphate estimates are not. The
informative signal is the within-pH phosphate-versus-non-phosphate contrast at matched pH; the pH~$6$
and pH~$8$ absolute scores are exploratory relative to the pH~$\approx7.3$ comparator reference.}
\label{tab:fumarase-kin}
\begin{tabular}{lccccl}
\toprule
Buffer (pH) & $\Kkin$ & $\Kthermo$ & $x$ & $\CH$ & Fold band \\
\midrule
phosphate (6.0) & 6.4 & 4.2 & 1.52 & 0.090 & within twofold \\
phosphate (7.0) & 6.1 & 4.2 & 1.45 & 0.070 & within twofold \\
phosphate (8.0) & 4.6 & 4.2 & 1.10 & 0.004 & within twofold \\
MES (6.0)       & 16  & 4.2 & 3.81 & 1.04  & 2--5-fold \\
TES (7.0)       & 19  & 4.2 & 4.52 & 1.37  & 2--5-fold \\
Tris (8.0)      & 17  & 4.2 & 4.05 & 1.15  & 2--5-fold \\
\bottomrule
\end{tabular}
\end{table}

\subsection{Triosephosphate isomerase and racemase controls}
\label{sec:racemases}

Triosephosphate isomerase pairs Putman and co-workers' bidirectional specificity
constants~\cite{Putman1972} ($\Kkin=20.4$) against Veech and co-workers' directly measured
equilibrium~\cite{Veech1969} ($\Kthermo=22.0$), for $x=0.93$ and $\CH=0.003$. The seven racemase
controls use the symmetry reference $\Kthermo=1$, so $\CH=\Jcost(\Kkin)$ tests only whether the forward and
reverse specificity constants are reciprocally consistent. Across all seven, $\CH$ ranges from $0$ to
$0.235$ (Table~\ref{tab:scored}); the largest, the \emph{B.\ anthracis} RacE1 control ($x=0.511$,
$\CH=0.235$), sits just below the twofold cut but is consistent within its (large) propagated
uncertainty, the underlying specificity constants carrying an independence-based spread
$\sigma_{\ln x}\approx0.22$ that already spans the boundary. A $\Kthermo=1$ control thus preserves
visible kinetic asymmetry instead of rounding it away. Two of the seven controls
(\emph{L.\ fermenti} glutamate racemase and \emph{P.\ aeruginosa} DadX) are reproduced or
database-extracted rather than re-extracted from primary tables, and serve as provenance-flagged
checks rather than primary two-sided tests.

\begin{table}[!htbp]
\centering
\caption{Representative scored records from the demonstration set. $\Kkin$ is the Haldane combination of matched
directional specificity terms from a single kinetic study; $\Kthermo$ is an independent thermodynamic
value ($=1$ for racemases under the usual biochemical convention); $x=\Kkin/\Kthermo$,
$\delta=\ln x$ is the signed log-ratio, and $\CH=\Jcost(x)$. Every row falls within the twofold
reporting band. The $\sigma_{\ln x,\mathrm{ind}}$ column gives the independence-based one-standard-deviation
spread of $\delta$ where a meaningful one is available (RacE1).}
\label{tab:scored}
\scriptsize
\setlength{\tabcolsep}{2.5pt}
\begin{tabular}{@{}>{\raggedright\arraybackslash}p{3.0cm}ccccccl@{}}
\toprule
Reaction (conditions) & $\Kkin$ & $\Kthermo$ & $x$ & $\delta=\ln x$ & $\CH$ & $\sigma_{\ln x,\mathrm{ind}}$ & Sources \\
\midrule
PGI, G6P $\rightleftharpoons$ F6P ($293$\,K) & 0.307 & 0.260 & 1.18 & $+0.17$ & 0.014 & --- & \cite{Stodeman2004,GoldbergTewari1995Iso} \\
TPI, GAP $\rightleftharpoons$ DHAP ($303$\,K kin.; $311$\,K thermo.)$^{b}$ & 20.4 & 22.0 & 0.93 & $-0.07$ & 0.003 & --- & \cite{Putman1972,Veech1969} \\
Mandelate racemase ($298$\,K) & 1.14 & 1.00 & 1.14 & $+0.13$ & 0.009 & --- & \cite{StMaurice2002} \\
Proline racemase ($T$ n.r.) & 1.00 & 1.00 & 1.00 & $0.00$ & 0.000 & --- & \cite{Fisher1986} \\
\emph{B. anthracis} RacE2 ($298$\,K) & 1.28 & 1.00 & 1.28 & $+0.25$ & 0.031 & --- & \cite{May2007} \\
\emph{B. anthracis} RacE1 ($298$\,K) & 0.511 & 1.00 & 0.511 & $-0.67$ & 0.235 & $0.22$ & \cite{May2007} \\
\emph{L. fermenti} GluR$^{\dagger}$ & 0.799 & 1.00 & 0.799 & $-0.22$ & 0.025 & --- & \cite{Gallo1993,May2007} \\
\emph{B. bifidum} AspR ($303$\,K) & 0.974 & 1.00 & 0.974 & $-0.03$ & $<0.001$ & --- & \cite{Yamashita2004} \\
\emph{P. aeruginosa} DadX$^{\dagger}$ ($296$\,K; $V_{\max}/K_M$) & 1.16 & 1.00 & 1.16 & $+0.15$ & 0.011 & --- & \cite{Strych2000} \\
\bottomrule
\end{tabular}
\begin{minipage}{\linewidth}
\vspace{3pt}\footnotesize
\emph{Notes.} $^{\dagger}$Provenance-limited control: numerical constants reproduced from a comparative
source or database extraction rather than independently verified from the primary kinetic table; these rows
serve as reciprocal-symmetry checks rather than condition-matched thermodynamic references.
$^{b}$The kinetic (303\,K) and thermodynamic (311\,K) measurements differ by 8\,K; the resulting
van't Hoff shift on the score is negligible at this level of discrepancy.
n.r.~=~not reported.
\end{minipage}
\end{table}

\begin{figure}[!htbp]
\centering
\includegraphics[width=0.85\textwidth]{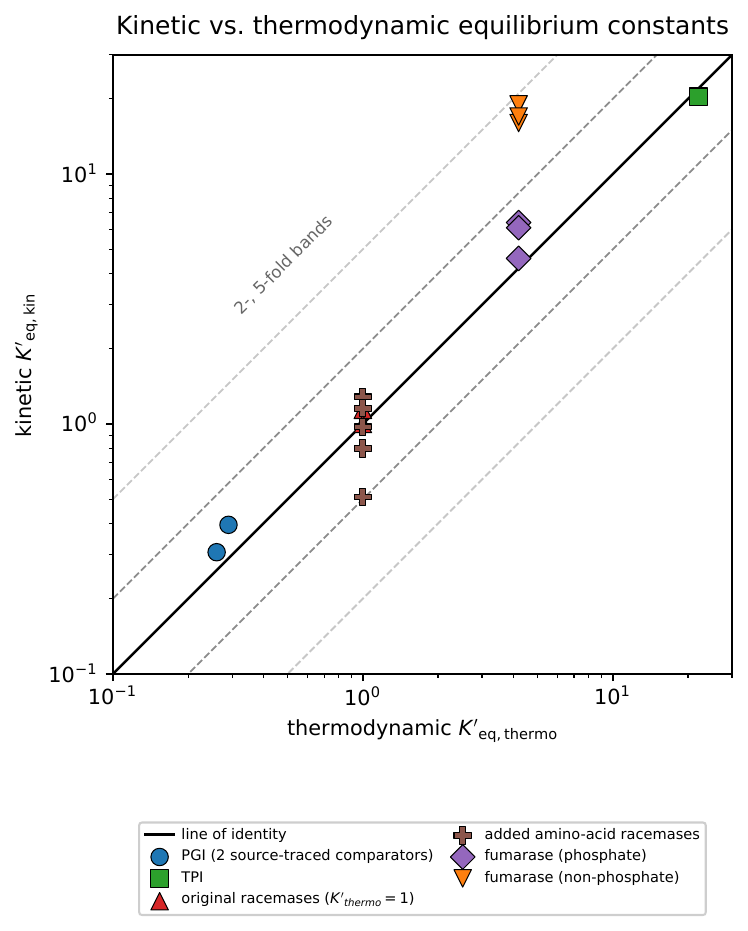}
\caption{Kinetic versus thermodynamic equilibrium constants for the demonstration-set comparisons
(Tables~\ref{tab:pgi},~\ref{tab:fumarase-kin}, and~\ref{tab:scored}). Axes are logarithmic; the bold
solid line is identity and dashed/dotted guides mark twofold and fivefold disagreement. The isomerases,
racemases (at $\Kthermo=1$ by enantiomer symmetry), and fumarase in phosphate buffer fall within
twofold; the non-phosphate fumarase estimates fall between the twofold and fivefold lines, their
absolute offsets conditional on the effective-uni--uni approximation.}
\label{fig:poc-consistency}
\end{figure}

\begin{figure}[!htbp]
\centering
\includegraphics[width=0.82\textwidth]{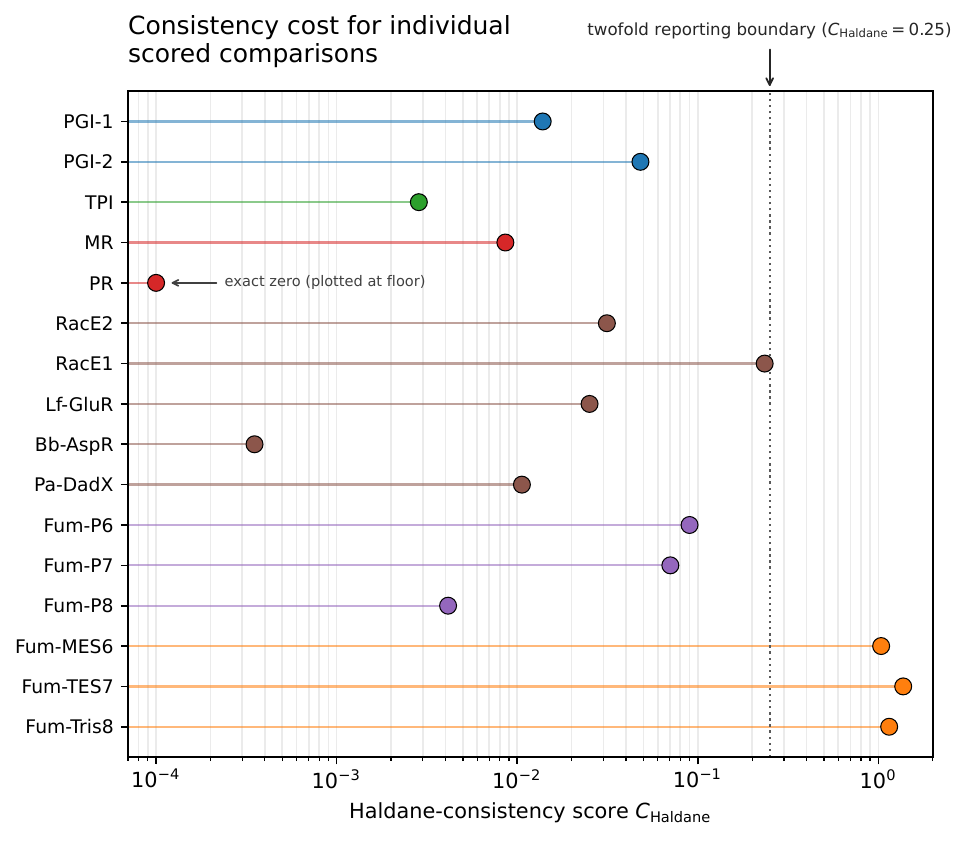}
\caption{Haldane-consistency score $\CH=\cosh(\ln x)-1$ for every demonstration-set comparison, on
a logarithmic axis, with the twofold reporting boundary ($\CH=0.25$) marked. The well-matched
comparisons fall well short of the boundary; the three non-phosphate fumarase rows ($\CH=1.0$--$1.4$)
exceed it under the effective-uni--uni approximation.}
\label{fig:poc-distribution}
\end{figure}

Figures~\ref{fig:poc-consistency} and~\ref{fig:poc-distribution} therefore serve different purposes. The
kinetic-versus-thermodynamic plot shows that the isomerase and racemase controls cluster near identity,
whereas the score distribution makes the fumarase buffer contrast visible on the same reciprocal scale;
together they motivate the twofold boundary used descriptively in the backbone and benchmark analyses.

\subsection{Mechanism-specific scoring for a bi--bi reaction}
\label{sec:poc-bibi}

Because the cost acts only on the ratio $\Kkin/\Kthermo$, extending the audit to a multi-substrate
reaction changes only the expression used to compute $\Kkin$; the score, its axioms, and its
interpretation are untouched. We make this concrete with wild-type horse-liver alcohol dehydrogenase
(EC 1.1.1.1),
ethanol $+$ NAD$^{+}\rightleftharpoons$ acetaldehyde $+$ NADH $+$ H$^{+}$, a compulsory-ordered bi--bi
mechanism in which the coenzyme binds first and dissociates last. This is an illustrative literature
compilation (Supplementary Table~S6), distinct from the audited yeast ADH-I backbone record in
Table~\ref{tab:backbone}. The mechanism-specific Haldane relation
is
\begin{equation}
\label{eq:haldane-bibi}
\Kkin=\frac{\kcatf}{\kcatr}\,\frac{K_{i,Q}\,K_{M,P}}{K_{i,A}\,K_{M,B}},
\end{equation}
with $A=\mathrm{NAD^{+}}$, $B=\mathrm{ethanol}$, $P=\mathrm{acetaldehyde}$, $Q=\mathrm{NADH}$, and
$K_{i,A},K_{i,Q}$ the coenzyme inhibition (dissociation) constants required by the thermodynamic
constraint. Substituting a single literature constant set (Supplementary Table~S6)
gives $\Kkin=1.6\times10^{-3}$ at the assay pH~8.0; against the evaluated thermodynamic value
($\Kthermo\approx6.8\times10^{-4}$ at pH~8.0~\cite{GoldbergOxidoreductases1993,Goldberg2004}) this is
$x\approx2.35$, $\CH\approx0.39$. This value is not classified as evidence of inconsistency: the six kinetic
constants are a literature compilation whose joint covariance is unavailable, and across plausible
correlations the standardized deviation stays a sub-$3\sigma$, unresolved discrepancy (Supplementary
Material). Two lessons transfer to scaling the audit. First, the workflow is unchanged by mechanism: only
$\Kkin$ is recomputed, and the score has the same interpretation as for a uni--uni record. Second, this is the
converse of a mechanism-mismatch audit --- applying the wrong (uni--uni) relation to such an enzyme can
manufacture a large, spurious $\CH$, the kind of assembly error represented by the benchmark surrogate in
Section~\ref{sec:benchmark}.

% ============================================================
\section{Audited two-sided backbone records}
\label{sec:backbone}

The demonstration set covers individual records rather than providing systematic breadth, and cannot establish how the score behaves across reactions. To evaluate the score on real, audited data we harvested a backbone of single-study
two-sided records under the prespecified protocol of Section~\ref{sec:design}, scoring each through the
appropriate uni--uni or mechanism-specific Haldane relation against a condition-matched comparator. The
limiting constraint throughout was the scarcity of single-study bidirectionality, not the score: most
reversible enzymes are characterized in only one direction. Successive harvest rounds --- open-access
primary literature, the offline BRENDA release, then a candidate-tracker expansion (under unchanged
acceptance criteria) into rare-sugar ketose epimerases, a TCA-cycle lyase, and the classical
steady-state oxidoreductase literature --- assembled twenty-one audited records, each passing the second
independent re-extraction audit. The full record-by-record harvest narrative, with the per-record constants,
orientations, comparator choices, and audit notes, is given in the Supplementary Material; here we report
the backbone composition and score distribution.

\paragraph{Mechanism coverage of the backbone}
Sixteen records are uni--uni and five are bi--bi, the latter spanning all three canonical
multi-substrate mechanisms. Table~\ref{tab:bibi} lists the mechanism-specific Haldane relations through
which the bi--bi records are scored, so that the exact expression assembling each apparent $\Kkin$ is in
the main text; $\CH=\Jcost(\Kkin/\Kthermo)$ is then formed exactly as for the uni--uni records. Each
expression is the zero-flux constraint of the corresponding reversible steady-state rate law in the
written orientation: the ordered forms carry the coenzyme dissociation constants required by the binding
order, the rapid-equilibrium random form uses dissociation constants because the binding steps are
equilibrated, and the ping-pong form uses the substituted-enzyme Haldane with a squared
limiting-velocity ratio. Applying the wrong (uni--uni) relation to such records is the kind of formula
assembly error represented by the mechanism-mismatch surrogate in Section~\ref{sec:benchmark}.
Published joint covariance matrices are not available for these five bi--bi records; their entries in
the backbone are therefore central-value Haldane audits with the same correlation-bracket treatment used
for the illustrative horse-liver ADH calculation. Unlike the horse-liver literature compilation, however,
all five backbone bi--bi records lie within twofold and are not used as evidence for an inconsistency.
The published sources also do not provide a common set of per-constant standard errors for all five rows,
so a main-text table of standardized log-deviations under both independence and a correlation bracket
would require imposing an error model not present in the primary literature. Where sufficient uncertainty
information is reported, the Supplementary Material gives the independence-based and bracketed
calculation; the backbone table therefore keeps these five rows as central-value audits. Because all five bi--bi
records lie within twofold, no backbone record's flagged/unflagged classification depends on the assumed
covariance: the five central-value bi--bi audits remain within twofold under any correlation, and the
three flagged records are uni--uni audits whose central fold discrepancies exceed twofold independently
of any covariance assumption. Reversing the bracket rule makes the same point as a sensitivity check:
flagging every central value above the twofold cut regardless of covariance would leave the backbone
count unchanged at three, because all five bi--bi records are within twofold. Only the illustrative
horse-liver ADH compilation (fold ${\approx}2.35$, not a backbone record) would change status, so the
flagged count is not an artifact of the covariance-bracketing discipline.

\begin{table}[!htbp]
\centering
\footnotesize
\setlength{\tabcolsep}{4pt}
\caption{Mechanism-specific Haldane relations used to score the five bi--bi backbone records. These
standard relations are the zero-flux constraints of the corresponding reversible steady-state rate laws
in the written orientation. Velocity subscripts follow each source's notation in the written (forward)
orientation; only the forward/reverse limiting-velocity ratio enters. The remaining constants are
$K_{ia},K_{iq}$ (coenzyme dissociation constants); $K_b,K_p$ (Michaelis constants of the second-bound
substrate and first-released product; for the rapid-equilibrium random mechanism $K_M=K_d$); and
$K_{PR},K_{OL}$ (the ternary-complex constants of the lactate-dehydrogenase form). In the ping-pong
relation, $K_{\mathrm{OAA}},K_{\mathrm{glu}},K_{\mathrm{asp}},K_{2\text{-OG}}$ are the Michaelis constants
for oxaloacetate (OAA), L-glutamate, L-aspartate, and 2-oxoglutarate (2-OG). The H$^{+}$ column
gives protons released in the written forward direction; for the proton-coupled rows the displayed
formulas are already expressed as apparent constants at the assay pH, so the corresponding
stoichiometric pH factor is absorbed into $\Kkin$ and $\Kthermo$. The mechanism changes only the
assembly of $\Kkin$; the score remains $\CH=\Jcost(\Kkin/\Kthermo)$ for every row. Full per-record
constants and primary sources are in Supplementary Table~S7.}
\label{tab:bibi}
\begin{tabular}{@{}>{\raggedright\arraybackslash}p{0.150\textwidth}>{\raggedright\arraybackslash}p{0.110\textwidth}>{\raggedright\arraybackslash}p{0.280\textwidth}>{\centering\arraybackslash}p{0.265\textwidth}c@{}}
\toprule
Mechanism & Record(s) & Reaction orientation & $\Kkin$ (Haldane) & H$^{+}$ \\
\midrule
Ordered bi--bi & ADH (yeast); MDH (pig heart) & ethanol $+$ NAD$^{+}$ $\rightleftharpoons$ acetaldehyde $+$ NADH $+$ H$^{+}$;\ \ (S)-malate $+$ NAD$^{+}$ $\rightleftharpoons$ OAA $+$ NADH $+$ H$^{+}$ & $\dfrac{V_f}{V_r}\dfrac{K_{iq}K_p}{K_{ia}K_b}$ & $+1$ \\[1.4ex]
Ordered bi--bi (ternary-complex form) & LDH (rabbit muscle) & (S)-lactate $+$ NAD$^{+}$ $\rightleftharpoons$ pyruvate $+$ NADH $+$ H$^{+}$ & $\dfrac{V_f\,K_{PR}}{V_r\,K_{OL}}$ & $+1$ \\[1.4ex]
Rapid-equilibrium random bi--bi & CK (rabbit muscle) & MgATP $+$ creatine $\rightleftharpoons$ MgADP $+$ phosphocreatine $+$ H$^{+}$ & $\dfrac{V_2}{V_1}\dfrac{K_{ia}K_b}{K_{iq}K_p}$$^{\ast}$ & $+1$ \\[1.4ex]
Ping-pong (substituted enzyme) & AAT (pig heart) & L-aspartate $+$ 2-oxoglutarate $\rightleftharpoons$ OAA $+$ L-glutamate & $\left(\dfrac{V_f}{V_r}\right)^{\!2}\dfrac{K_{\mathrm{OAA}}K_{\mathrm{glu}}}{K_{\mathrm{asp}}K_{2\text{-OG}}}$ & $0$ \\
\bottomrule
\end{tabular}
\begin{minipage}{\linewidth}
\vspace{3pt}\footnotesize
$^{\ast}$In Morrison and James's~\cite{MorrisonJames1965} notation for creatine kinase, $V_2$ is the written-forward
limiting velocity (phosphocreatine formation) and $V_1$ is the reverse; $V_2<V_1$ is consistent with
the equilibrium favoring the reactants at standard conditions.
\end{minipage}
\end{table}

\paragraph{Derivation of the mechanism-specific relations}
Each entry of Table~\ref{tab:bibi} follows by setting the numerator of the reversible steady-state rate
law to zero in the written direction. For an ordered bi--bi mechanism with first-substrate and
first-product (coenzyme) dissociation constants $K_{ia}$ and $K_{iq}$, the zero-flux balance is
\begin{equation}
\label{eq:bibi-zeroflux}
\frac{V_f}{K_{ia}K_b}\,[A][B]=\frac{V_r}{K_{iq}K_p}\,[P][Q],
\qquad\text{giving}\qquad
\Kkin=\frac{[P][Q]}{[A][B]}=\frac{V_f}{V_r}\,\frac{K_{iq}K_p}{K_{ia}K_b}.
\end{equation}
For the rapid-equilibrium random creatine-kinase form, Morrison and James's notation gives the zero-flux
balance
\[
\frac{V_2}{K_{iq}K_p}[\mathrm{MgATP}][\mathrm{creatine}]
=
\frac{V_1}{K_{ia}K_b}[\mathrm{MgADP}][\mathrm{phosphocreatine}],
\]
so $\Kkin=(V_2/V_1)(K_{ia}K_b)/(K_{iq}K_p)$ in the written direction. For lactate dehydrogenase, Hakala, Glaid, and Schwert's ternary-complex
kinetic form~\cite{Hakala1956} has numerator
$(V_f/K_{OL})[\mathrm{lactate}][\mathrm{NAD}^{+}]-(V_r/K_{PR})[\mathrm{pyruvate}][\mathrm{NADH}]$,
giving $\Kkin=V_fK_{PR}/(V_rK_{OL})$. For the substituted-enzyme ping-pong aminotransferase, the
zero-flux numerator contains the two half-reaction velocity ratios and reduces, in the written direction,
to
\[
\Kkin=\left(\frac{V_f}{V_r}\right)^2
\frac{K_{\mathrm{OAA}}K_{\mathrm{glu}}}{K_{\mathrm{asp}}K_{2\text{-OG}}}.
\]
These mechanism-specific balances show where the extra constants enter; the score itself is unchanged.

\begin{figure}[!htbp]
\centering
\includegraphics[width=0.95\textwidth]{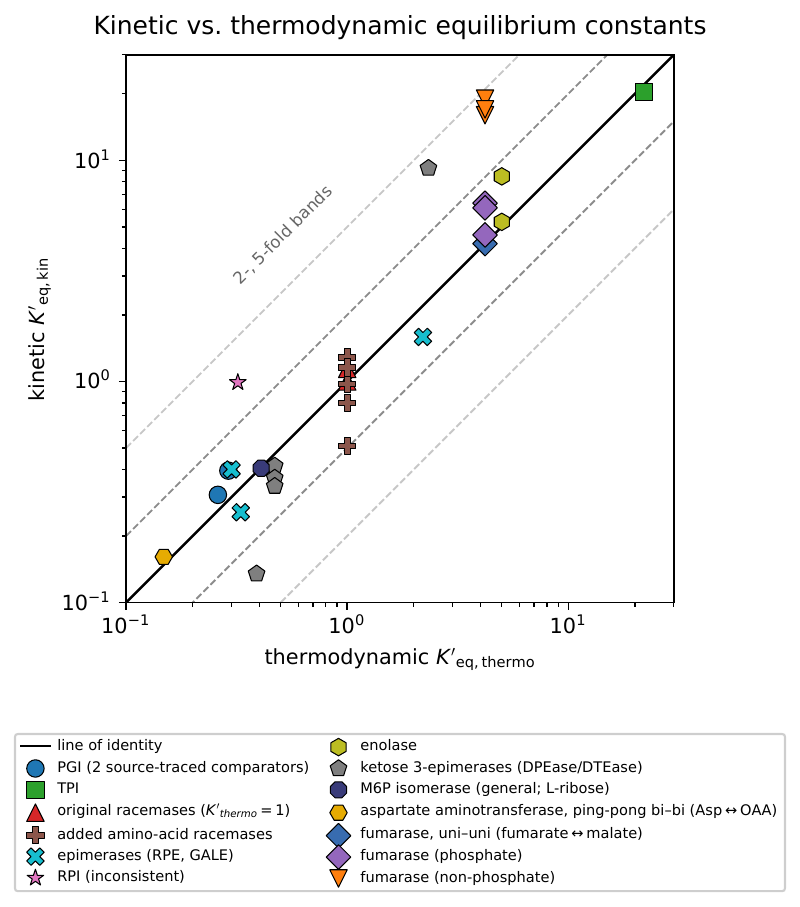}
\caption{Kinetic versus thermodynamic apparent equilibrium constants for the central-range backbone
records (sixteen of the twenty-one) and the demonstration-only within-twofold seed records
(the eleven seeds not already in the backbone: racemases clustered at the unit-equilibrium diagonal
and the phosphate-buffer fumarase rows). Axes are logarithmic;
the solid line is identity and the dashed/dotted guides mark twofold and fivefold disagreement. Pea
phosphoriboisomerase, the \emph{C.\ scindens} D-psicose 3-epimerase, and the \emph{Sinorhizobium}
D-tagatose 3-epimerase fall in the 2--5-fold region (the three flagged records); the remainder sit
within twofold. The five lowest-$\Kthermo$ records fall below the plotted range and appear in the
complete score summary of Figure~\ref{fig:backbone-distribution}.}
\label{fig:backbone-consistency}
\end{figure}

\begin{figure}[!htbp]
\centering
\includegraphics[width=0.62\textwidth]{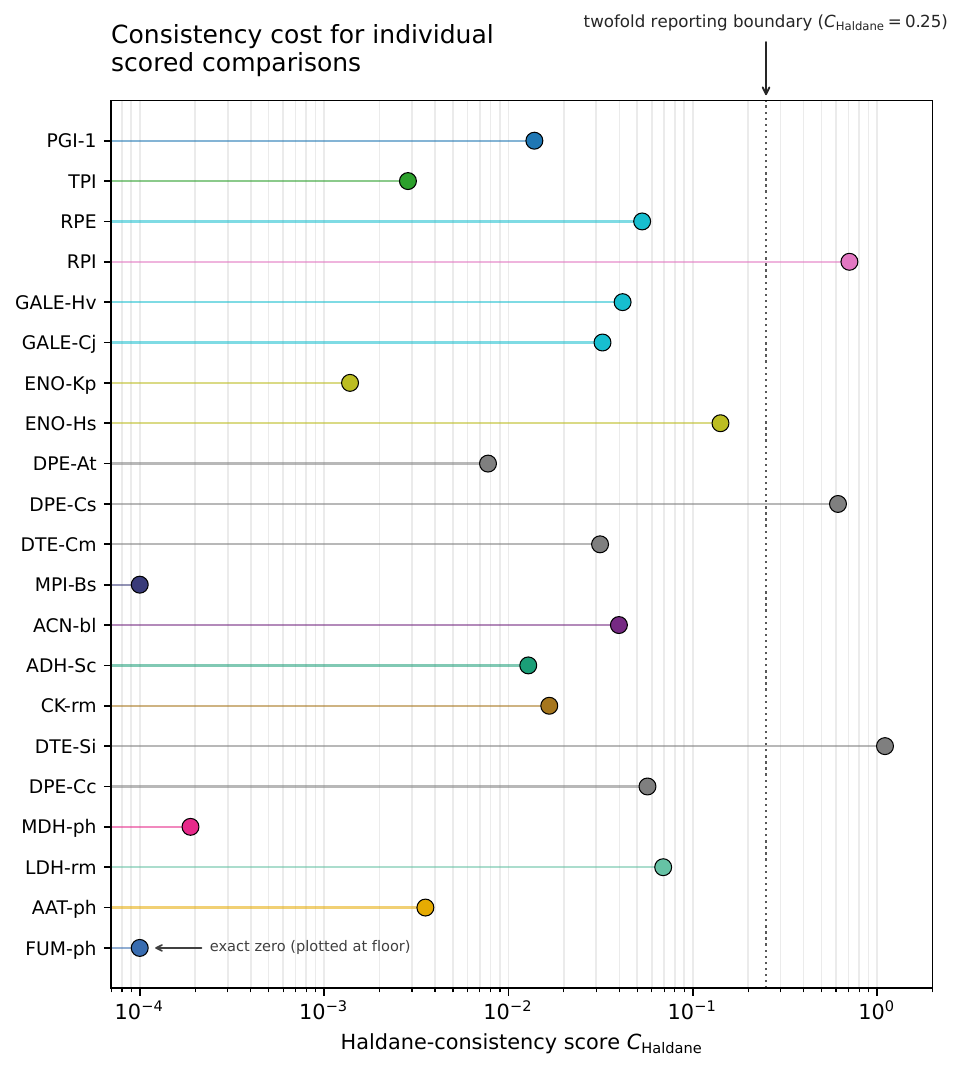}
\caption{Haldane-consistency score $\CH=\cosh(\ln x)-1$ for all twenty-one audited real two-sided
records (identified in Table~\ref{tab:backbone}), on a logarithmic axis with the twofold reporting
boundary ($0.25$) marked. All records are scored exactly as the demonstration-set records, and only three
--- pea phosphoriboisomerase (RPI), the \emph{C.\ scindens} D-psicose 3-epimerase (DPE-Cs), and the
\emph{Sinorhizobium} D-tagatose 3-epimerase (DTE-Si) --- exceed the twofold boundary.}
\label{fig:backbone-distribution}
\end{figure}

\paragraph{Backbone score distribution}
The twenty-one records are summarized in Table~\ref{tab:backbone}.
Figure~\ref{fig:backbone-consistency} places the central-range records on the
kinetic-versus-thermodynamic plane, and Figure~\ref{fig:backbone-distribution} shows the full score
distribution. Eighteen fall within the twofold band and three are flagged inconsistencies spanning three
distinct reactions: pea phosphoriboisomerase, the \emph{C.\ scindens} D-psicose 3-epimerase, and the
\emph{Sinorhizobium} D-tagatose 3-epimerase. All three flagged records are carbohydrate isomerases or
epimerases (EC~5.3.1 and 5.1.3), the class in which the backbone is chemically concentrated, so the
flagged tail is a within-family observation for carbohydrate isomerase/epimerase chemistry rather than a
general enzyme-kinetics claim. The count clears the fifteen-record inclusion threshold
with wide margin, so the backbone supplies a descriptive empirical reference distribution for this
curated sample rather than a population-level estimate of enzyme-kinetics consistency. Eight records are genuinely independent two-sided tests, with kinetics fit under no
thermodynamic prior and scored against separately measured equilibrium constants (the five bi--bi
records, pig-heart fumarase, PGI, and TPI); a further five records carry
same-study comparators --- the tightest possible condition match --- four of them within twofold. One
record (human-muscle enolase) is comparator-sensitive in the sense made precise in Section~\ref{sec:discussion}.

{\footnotesize
\setlength{\tabcolsep}{3pt}
\setlength{\LTcapwidth}{\textwidth}
\begin{longtable}{@{}>{\raggedright\arraybackslash}p{0.245\textwidth}c c c c c c >{\raggedright\arraybackslash}p{0.13\textwidth}@{}}
\caption{The twenty-one audited real two-sided records of the backbone (Section~\ref{sec:backbone};
the eight provenance-filtered independent tests are marked in the ``Ind.'' column),
each scored through the appropriate (uni--uni or, for the five bi--bi records, the mechanism-specific)
Haldane relation against a condition-matched thermodynamic comparator. $\Kkin$ is the kinetic apparent
equilibrium constant, $\Kthermo$ the comparator, fold $=\max(x,1/x)$ with $x=\Kkin/\Kthermo$, $\CH$ the
consistency score~\eqref{eq:CHdef}, and the band is the descriptive fold-discrepancy category of
Table~\ref{tab:bands}. Eighteen are within twofold; three (RPI, the \emph{C.\ scindens} epimerase, and
the \emph{Sinorhizobium} D-tagatose 3-epimerase) are flagged inconsistencies. Per-record constants,
mechanism-specific bi--bi forms, scores, and primary sources are in Supplementary Tables~S7--S9.}
\label{tab:backbone}\\
\toprule
Enzyme (organism) & Ind. & EC & $\Kkin$ & $\Kthermo$ & fold & $\CH$ & Band \\
\midrule
\endfirsthead
\multicolumn{8}{@{}l}{\footnotesize\emph{Table~\ref{tab:backbone} (continued).}}\\
\toprule
Enzyme (organism) & Ind. & EC & $\Kkin$ & $\Kthermo$ & fold & $\CH$ & Band \\
\midrule
\endhead
\midrule
\multicolumn{8}{r@{}}{\footnotesize\emph{(continued on next page)}}\\
\endfoot
\bottomrule
\endlastfoot
RPE, CrRPE1 (\emph{C.\ reinhardtii})    & -- & 5.1.3.1  & 1.59 & 2.2  & 1.38 & 0.053 & within twofold \\
RPI (\emph{P.\ sativum}, pea)           & -- & 5.3.1.6  & 0.99 & 0.32 & 3.09 & 0.71  & 2--5-fold \\
GALE, HvUGE1 (\emph{H.\ vulgare})       & -- & 5.1.3.2  & 0.40 & 0.30 & 1.33 & 0.042 & within twofold \\
GALE, Gne (\emph{C.\ jejuni})           & -- & 5.1.3.2  & 0.26 & 0.33 & 1.29 & 0.032 & within twofold \\
Enolase (\emph{K.\ pneumoniae})         & -- & 4.2.1.11 & 5.27 & 5.0  & 1.05 & 0.001 & within twofold \\
Enolase (human muscle)$^{a}$            & -- & 4.2.1.11 & 8.46 & 5.0  & 1.69 & 0.14  & within twofold \\
DPEase (\emph{A.\ tumefaciens})         & -- & 5.1.3.30 & 0.42 & 0.47 & 1.13 & 0.008 & within twofold \\
DPEase (\emph{C.\ scindens})            & -- & 5.1.3.30 & 0.14 & 0.39 & 2.88 & 0.61  & 2--5-fold \\
DPEase (\emph{C.\ cellulolyticum})      & -- & 5.1.3.30 & 0.34 & 0.47 & 1.40 & 0.057 & within twofold \\
DTEase (\emph{C.\ minuta})              & -- & 5.1.3.31 & 0.37 & 0.47 & 1.28 & 0.031 & within twofold \\
DTEase, sDTE (\emph{Sinorhizobium})$^{b}$ & -- & 5.1.3.31 & 9.22 & 2.33 & 3.96 & 1.10 & 2--5-fold \\
M6P isomerase (\emph{B.\ subtilis})     & -- & 5.3.1.8  & 0.41 & 0.41 & 1.01 & 0.000 & within twofold \\
Aconitase (beef liver)                  & -- & 4.2.1.3  & 0.102 & 0.077 & 1.32 & 0.040 & within twofold \\
Alcohol DH (yeast, WT)$^{c}$            & yes & 1.1.1.1  & $2.3\!\times\!10^{-4}$ & $2.0\!\times\!10^{-4}$ & 1.17 & 0.013 & within twofold \\
Creatine kinase (rabbit muscle)$^{d}$   & yes & 2.7.3.2  & 0.036 & 0.030 & 1.20 & 0.017 & within twofold \\
Malate DH (pig heart)$^{e}$             & yes & 1.1.1.37 & $1.04\!\times\!10^{-4}$ & $1.02\!\times\!10^{-4}$ & 1.02 & 0.0002 & within twofold \\
Lactate DH (rabbit muscle)$^{f}$        & yes & 1.1.1.27 & $1.81\!\times\!10^{-5}$ & $2.62\!\times\!10^{-5}$ & 1.45 & 0.069 & within twofold \\
Aspartate aminotransferase (pig heart)$^{g}$ & yes & 2.6.1.1 & 0.161 & 0.148 & 1.09 & 0.003 & within twofold \\
Fumarase (pig heart)$^{h}$              & yes & 4.2.1.2 & 4.20 & 4.20 & 1.00 & 0.000 & within twofold \\
Phosphoglucose isomerase (yeast)$^{i}$  & yes & 5.3.1.9 & 0.307 & 0.260 & 1.18 & 0.014 & within twofold \\
Triosephosphate isomerase (chicken)$^{i,j}$ & yes & 5.3.1.1 & 20.4 & 22.0 & 1.08 & 0.003 & within twofold \\
\end{longtable}
\par}

\vspace{-2pt}
\noindent{\footnotesize\emph{Notes.}~Enzyme abbreviations: RPE, ribulose-5-phosphate 3-epimerase; RPI,
ribose-5-phosphate isomerase (phosphoriboisomerase); GALE, UDP-glucose 4-epimerase; DPEase, D-psicose
3-epimerase; DTEase, D-tagatose 3-epimerase; M6P, mannose-6-phosphate. $\Kkin$ and $\Kthermo$ are
displayed to three significant figures;
$\CH$ and the fold error are computed from full-precision values (Supplementary Table~S8). ``Ind.'' marks records whose kinetic constants were fit without a thermodynamic prior and scored against separately measured equilibria.
$^{a}$\,Human-muscle enolase is comparator-sensitive (within twofold against the condition-matched
$K_{\mathrm{app}}\approx5$; $2$--$5$-fold against the standard-state $\approx3.6$;
Section~\ref{sec:discussion}). $^{b}$\,\emph{Sinorhizobium} sDTE is scored on its native
D-tagatose$\rightleftharpoons$D-sorbose pair against the organism-independent $\approx30{:}70$
equilibrium, robustly inconsistent across the comparator range.
$^{c,d,e,f,g}$\,The five bi--bi records, scored through the
mechanism-specific Haldane relations of Table~\ref{tab:bibi} (ADH, MDH, LDH ordered; CK
rapid-equilibrium random; AAT ping-pong). $^{h}$\,Fumarase, the original effective
uni--uni Haldane test of Bock and Alberty~\cite{BockAlberty1953}; this pig-heart backbone record is
distinct from the \emph{C.\ glutamicum} demonstration case of Section~\ref{sec:fumarase}. $^{i}$\,The two demonstration-set isomerases promoted into
the backbone. The five bi--bi records together with fumarase and the two isomerases are the eight
genuinely independent tests (kinetics and equilibrium taken from separate studies).
$^{j}$\,The TPI kinetic constants (Putman et al.\ 1972) were measured at \SI{303}{\kelvin} while the
thermodynamic comparator (Veech et al.\ 1969) was determined at \SI{311}{\kelvin}; the resulting
van't Hoff shift on the score is negligible at this level of discrepancy (see
Table~\ref{tab:scored}, note~$^{b}$). The comparator is an in vivo apparent equilibrium measured in
rat liver~\cite{Veech1969}; the organism difference is not expected to affect the reaction
thermodynamics.\par}

\begin{table}[!htbp]
\centering
\caption{Backbone sensitivity summaries. The independent-test subset pairs kinetic constants fit without
a thermodynamic prior against separately measured equilibrium constants (the five bi--bi records,
pig-heart fumarase, PGI, and TPI). The pre-expansion subset removes the rare-sugar DPEase/DTEase records
added after the candidate tracker was widened, but applies the same inclusion criteria. All values are
computed from the same full-precision records used in Table~\ref{tab:backbone}. With $n=8$, the
independent subset is the primary provenance subset for downstream empirical summaries, but its median
and maximum are order-of-magnitude descriptors rather than confidence-estimated distributional
parameters. The exact one-sided $95\%$ Clopper--Pearson upper bound on the underlying beyond-twofold
rate for the observed $0/8$ exceedances is $1-0.05^{1/8}\approx0.31$, shown in the table so that the
binomial uncertainty of the $8/8$ result is visible at the record level rather than only in the text.}
\label{tab:independent-subset}
\footnotesize
\begin{tabular}{@{}>{\raggedright\arraybackslash}p{0.30\textwidth}
                >{\centering\arraybackslash}p{0.06\textwidth}
                >{\raggedright\arraybackslash}p{0.54\textwidth}@{}}
\toprule
Subset & $n$ & Summary \\
\midrule
Independent tests only & 8 & 8/8 within $2\times$; flagged 0; median $\CH=0.008$; maximum $\CH=0.069$ (LDH); one-sided $95\%$ upper bound on the beyond-twofold rate ${\approx}0.31$ (Clopper--Pearson) \\
Pre-expansion tracker subset & 16 & 15/16 within $2\times$; flagged 1; median $\CH=0.016$; maximum $\CH=0.71$ (RPI) \\
\bottomrule
\end{tabular}
\end{table}

Table~\ref{tab:independent-subset} shows that the independent subset is uniformly within twofold
(fold errors $1.00$--$1.45$). With only eight records, these summaries are descriptive checks on scale,
not distributional estimates. The tracker-expansion sensitivity leaves one flagged record in sixteen;
the two additional flags therefore enter with the rare-sugar DPEase/DTEase expansion, whereas the
near-zero independent subset does not.

\FloatBarrier

Together, Table~\ref{tab:backbone} and Figures~\ref{fig:backbone-consistency}--\ref{fig:backbone-distribution}
show that the backbone is not merely a larger version of the demonstration set. It supplies a
threshold-cleared descriptive reference distribution, includes mechanism-specific bi--bi records, and
separates the three flagged records from the within-twofold majority without overlap at the twofold
boundary.

% ============================================================
\section{Semi-synthetic benchmark}
\label{sec:benchmark}

The real backbone supplies an empirical reference distribution but no ground-truth error labels: with a
handful of author-acknowledged exceptions, the error status of individual real records is unknown. To
estimate how the fold-band classification performs we therefore build a semi-synthetic labeled benchmark.
Because the score is monotone in $|\ln x|$, every operating characteristic reported in this section ---
the area under the ROC curve and the per-cut sensitivities alike --- is a diagnostic of the injected
error taxonomy and its detectability at the fixed cuts, not a performance metric peculiar to $\CH$.
Beginning from the within-twofold seed records (labeled consistent), we inject biochemically
realistic, individually known error modes to produce labeled inconsistent cases:
\begin{enumerate}
  \item a direction swap, $\Kkin\mapsto1/\Kkin$, modeling reverse constants reported as forward;
  \item a unit transcription error (micromolar versus millimolar), a factor of $10^{3}$;
  \item an omitted or wrong proton correction across a unit pH step, a factor of $10^{\pm1}$;
  \item a condition mismatch with the comparator effectively taken at the wrong temperature, the
  ratio offset fixed by the van't Hoff relation~\cite{Goldberg2004};
  \item an isoform/organism mismatch, drawn as a moderate log-normal pairing error
  ($\sigma=\ln2$); and
  \item a mechanism/formula mismatch, represented by a broad log-normal surrogate for using the
  wrong Haldane assembly formula ($\sigma=\ln5$), the error type illustrated by Section~\ref{sec:poc-bibi}.
\end{enumerate}
Each seed contributes one unperturbed negative at its measured $\Kkin$, and each mode multiplies that
estimate by a mode-specific factor to yield a labeled positive carrying a known log-offset: the
deterministic direction swap; $10^{\pm3}$ and $10^{\pm1}$ with sign drawn uniformly; the van't Hoff
factor with reaction enthalpy $\Delta H\sim\mathcal{U}(-60,+60)~\mathrm{kJ\,mol^{-1}}$ and assay offset
$\sim\mathcal{U}(-12,+12)~\mathrm{K}$; and the two log-normal modes. The temperature-mismatch mode is
implemented as a deliberately broad stress-test prior rather than an empirical enthalpy distribution for
the carbohydrate-rich backbone, and Table~\ref{tab:val-param-sensitivity} reports the effect of halving
and doubling those ranges. The baseline log-normal widths encode a moderate twofold isoform/organism
pairing error ($\sigma=\ln2$) and a deliberately broad fivefold mechanism/formula surrogate
($\sigma=\ln5$), so their detectability is read as sensitivity to the chosen taxonomy rather than as an
empirical prevalence estimate. It is represented as the equivalent multiplicative displacement of
$x=\Kkin/\Kthermo$ through $\Kkin$; the ROC
results depend on the ratio displacement, not on which side of the comparison was mis-curated. Likewise,
the mechanism/formula mode is a log-normal surrogate applied to the ratio, not a literal reassembly of
mechanism-tagged bi--bi constants. We draw twenty replicates per mode per seed from a single fixed
pseudo-random stream (Mersenne Twister, seed $0$), de-duplicating identical deterministic rows, so that
the three stochastic modes retain twenty draws while the deterministic modes collapse to their distinct
values. The twenty-nine within-twofold seeds consist of eighteen within-twofold backbone records and
eleven demonstration-only seeds; the 60KAH PGI record and TPI appear in both displays and are counted once, whereas the 68DYS PGI record is a distinct demonstration-only seed. This
gives twenty-nine negatives and $1{,}885$ labeled positives.

\begin{figure}[!htbp]
\centering
\includegraphics[width=0.62\textwidth]{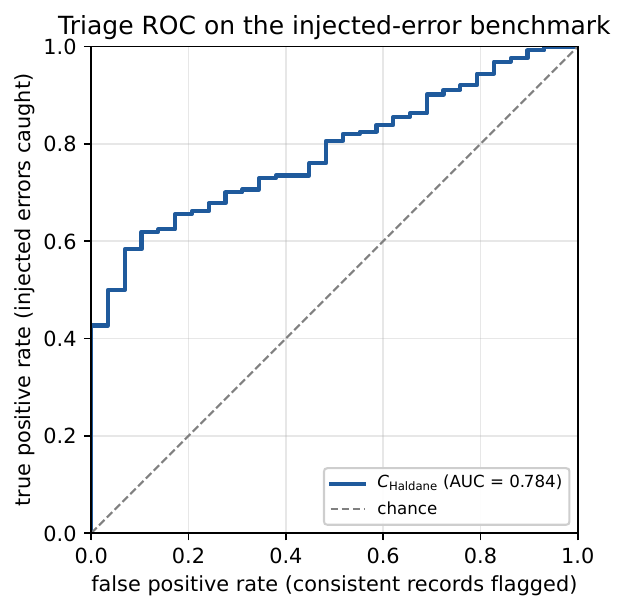}
\caption{Receiver-operating-characteristic curve for the Haldane-consistency score $\CH$ as a continuous
classifier on the injected-error benchmark built from the within-twofold seed set (twenty-nine seed
records, $1{,}885$ injected errors; area under the curve $0.784$, $95\%$ bootstrap CI
$0.725$--$0.838$). The curve and its area are conditional on the injected error taxonomy of
Section~\ref{sec:benchmark} and on the within-twofold seed set, and so characterize the workflow rather
than its
external performance.}
\label{fig:val-roc}
\end{figure}

\begin{figure}[!htbp]
\centering
\includegraphics[width=0.82\textwidth]{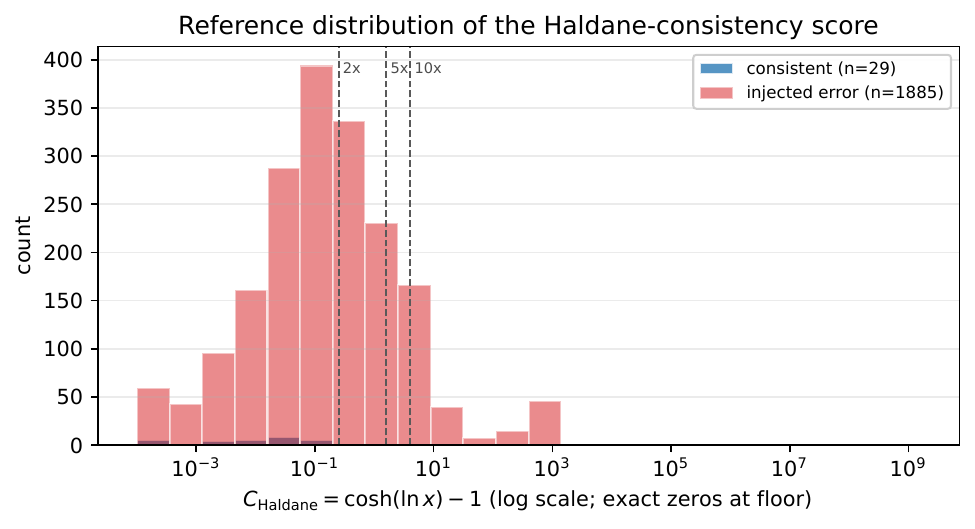}
\caption{Reference distribution of $\CH=\cosh(\ln x)-1$ for the within-twofold seed records (blue,
$n=29$) and the injected-error cases (red, $n=1{,}885$) on a logarithmic score axis, with the $2$-,
$5$-, and $10$-fold band boundaries marked. The seed records concentrate at small $\CH$ while injected
errors spread across and beyond the bands. The histogram is conditional on the injected error taxonomy
and the within-twofold seed set.}
\label{fig:val-reference}
\end{figure}

Treated as a continuous classifier, $\CH$ attains an area under the ROC curve (AUC) of $0.784$ ($95\%$
stratified-bootstrap CI $0.725$--$0.838$, $2{,}000$ resamples; Figure~\ref{fig:val-roc}). This AUC is
identical for any strictly monotone transformation of $|\ln x|$ --- including $|\ln x|$, $(\ln x)^2$,
or $|\Delta\Delta G|$ --- and therefore measures the discriminability of the fold-error information
rather than a performance advantage unique to $\CH$. The
within-twofold seed records pile up near $\CH=0$ while injected errors spread across the fold-band
boundaries (Figure~\ref{fig:val-reference}). At the fixed cuts, the relevant non-tautological quantity is
the fraction of injected errors that exceed each cut; this sensitivity falls as the band widens because
many injected offsets are deliberately small (Table~\ref{tab:val-oc}). The within-twofold seed set is the
negative class by construction, so its absence of flags is not an external false-positive estimate. The
per-mode breakdown (Table~\ref{tab:val-modes}) identifies the two least-detectable error
modes: a direction swap when $\Kkin\approx1$ (often alongside $\Kthermo=1$ in racemase controls) barely
moves the ratio, and a small temperature-driven condition mismatch is the hardest mode of all; gross unit
and proton errors, by contrast, are caught essentially always.

Two robustness checks confirm the area is not an artifact of how positives are counted
(Table~\ref{tab:val-sensitivity}). A seed-level cluster bootstrap that resamples the twenty-nine
seeds with their injected perturbations returns the same point estimate ($0.784$, CI $0.743$--$0.828$),
so clustering neither inflates the estimate nor widens the interval. And because the row-wise area
weights every injected case equally, it is dominated by the three log-normal modes and so understates
per-mode discrimination: weighting the six modes equally raises the area to $0.861$, and removing the
single hardest mode raises the row-wise area to $0.857$. The row-weighted AUC of $0.784$ is thus a deliberately
conservative, mixture-weighted summary, not a ceiling on the score's discrimination. Increasing the
stochastic replication budget from twenty to fifty or one hundred draws per mode changed the row-wise AUC
only modestly ($0.779$ and $0.771$, respectively) and left the equal-mode AUC essentially unchanged
($0.863$ and $0.861$). Narrowing the condition-mismatch mode toward a bibliographically anchored enthalpy
distribution --- approximated here by halving its van't Hoff enthalpy and temperature ranges
(Table~\ref{tab:val-param-sensitivity}) --- lowers the equal-mode AUC only from $0.861$ to $0.844$, so the
mixture-balanced summary does not depend on the deliberately broad stress-test prior.

These operating characteristics are conditional on the injected error taxonomy and on the within-twofold
seed set: they
describe the implemented workflow's response to those errors, not the full and unknown distribution of
external curation mistakes. Taken with the parameter perturbations in
Table~\ref{tab:val-param-sensitivity}, the row-wise AUC should be read as approximately
$0.78\pm0.05$ under plausible changes to the injected taxonomy and prior widths, rather than as an
intrinsic property of the score. The benchmark generator, the operating-characteristics analysis, and the
SHA-256 checksums of their regenerated outputs accompany the repository
(Data and Code Availability), so every numerical result here is exactly reproducible.

\begin{table}[!htbp]
\centering
\caption{Injected-error sensitivity of the fixed fold-band cuts on the semi-synthetic benchmark
(twenty-nine within-twofold seed records, $1{,}885$ injected errors; an injected case is detected when
its fold error exceeds the cut). These estimates are conditional on the injected error taxonomy of
Section~\ref{sec:benchmark} and the within-twofold seed set. Specificity, precision, and false-positive
rate are not tabulated because the negative class is defined by the within-twofold cut and would make
those quantities tautological rather than external false-positive estimates.}
\label{tab:val-oc}
\begin{tabular}{@{}>{\centering\arraybackslash}p{0.42\textwidth}>{\centering\arraybackslash}p{0.42\textwidth}@{}}
\toprule
Fold cut & Sensitivity \\
\midrule
$2\times$  & $0.42$ \\
$5\times$  & $0.19$ \\
$10\times$ & $0.11$ \\
\bottomrule
\end{tabular}
\end{table}

\begin{table}[!htbp]
\centering
\caption{Per-error-mode detectability at the twofold cut: the fraction of injected cases of each mode
whose fold error exceeds $2\times$. The near-symmetric direction swap and the small condition mismatch
are least detectable, while gross unit and
proton errors are always caught. For the isoform/organism log-normal mode with $\sigma=\ln2$, the
observed fraction ($0.34$) is close to the symmetric analytic expectation near one third, so that row
mainly checks calibration of the injected distribution. The mechanism/formula row is likewise a broad
log-normal surrogate whose detectability reflects the chosen $\sigma=\ln5$ as well as formula assembly
errors. Conditional on the injected taxonomy and the within-twofold seed set.}
\label{tab:val-modes}
\begin{tabular}{@{}>{\raggedright\arraybackslash}p{0.42\textwidth}>{\centering\arraybackslash}p{0.22\textwidth}>{\centering\arraybackslash}p{0.18\textwidth}@{}}
\toprule
Injected error mode & Detected\,/\,total & Fraction \\
\midrule
Unit transcription ($\times10^{3}$)            & $58/58$   & $1.00$ \\
Proton miscorrection ($\times10^{\pm1}$)       & $58/58$   & $1.00$ \\
Mechanism/formula mismatch (log-normal surrogate) & $386/580$ & $0.67$ \\
Direction swap ($\Kkin\mapsto1/\Kkin$)         & $22/29$   & $0.76$ \\
Isoform / organism mismatch                    & $200/580$ & $0.34$ \\
Condition mismatch (wrong temperature)         & $64/580$  & $0.11$ \\
\bottomrule
\end{tabular}
\end{table}

\begin{table}[!htbp]
\centering
\caption{Robustness of the area under the ROC curve to the injected-error mixture and to the clustering
of injected cases by seed (fixed seed-0 benchmark, published scorer). The row-wise value
weights every injected case equally and so is dominated by the three log-normal modes; the seed-level
cluster bootstrap resamples the twenty-nine seeds with their perturbations, and the mixture-balanced
value averages the per-mode areas. The per-mode areas show that the row-wise value is reduced by the two
hardest modes alone.}
\label{tab:val-sensitivity}
\begin{tabular}{@{}>{\raggedright\arraybackslash}p{0.56\textwidth}>{\centering\arraybackslash}p{0.34\textwidth}@{}}
\toprule
AUC summary & Value ($95\%$ CI) \\
\midrule
Row-wise (stratified bootstrap)                        & $0.784$ ($0.725$--$0.838$) \\
Seed-level cluster bootstrap                           & $0.784$ ($0.743$--$0.828$) \\
Equal-weight-per-mode (mixture-balanced)               & $0.861$ \\
Excluding the hardest mode (condition mismatch)        & $0.857$ \\
\midrule
\multicolumn{2}{@{}l}{\emph{Per-mode AUC (mode positives vs.\ the within-twofold seed records):}}\\
\quad unit transcription; proton miscorrection         & $1.00$;\ \ $1.00$ \\
\quad mechanism/formula mismatch (log-normal surrogate) & $0.905$ \\
\quad direction swap                                   & $0.863$ \\
\quad isoform / organism mismatch                      & $0.781$ \\
\quad condition mismatch (wrong temperature)           & $0.619$ \\
\bottomrule
\end{tabular}
\end{table}

\begin{table}[!htbp]
\centering
\caption{Sensitivity of the benchmark summaries to injected-error parameter choices. Variants were
regenerated with the same seed set, random seed, and scorer as Table~\ref{tab:val-sensitivity}. The
``narrow'' and ``wide'' rows vary the two log-normal error modes by $\pm\ln2$
($\sigma_{\mathrm{isoform}}=\ln2$ and $\sigma_{\mathrm{mechanism}}=\ln5$ at baseline). The van't Hoff
rows halve or double both the enthalpy and temperature ranges used for the condition-mismatch mode.}
\label{tab:val-param-sensitivity}
\footnotesize
\begin{tabular}{@{}>{\raggedright\arraybackslash}p{0.27\textwidth}
                >{\centering\arraybackslash}p{0.21\textwidth}
                >{\raggedright\arraybackslash}p{0.42\textwidth}@{}}
\toprule
Variant & AUCs & Detectability at $2\times$ cut \\
\midrule
Baseline & row $0.784$; equal-mode $0.861$ & mechanism $0.67$; isoform $0.34$; condition $0.11$ \\
Log-normal widths $-\ln2$ & row $0.744$; equal-mode $0.801$ & mechanism $0.47$; isoform $0.00$; condition $0.11$ \\
Log-normal widths $+\ln2$ & row $0.828$; equal-mode $0.885$ & mechanism $0.78$; isoform $0.62$; condition $0.11$ \\
Van't Hoff ranges halved & row $0.753$; equal-mode $0.844$ & mechanism $0.67$; isoform $0.34$; condition $0.02$ \\
Van't Hoff ranges doubled & row $0.841$; equal-mode $0.892$ & mechanism $0.67$; isoform $0.34$; condition $0.51$ \\
\bottomrule
\end{tabular}
\end{table}

The validation tables and figures connect the fixed fold bands to operating characteristics rather than
introducing new thresholds. Table~\ref{tab:val-oc} gives the injected-error sensitivity of the fixed cuts,
Table~\ref{tab:val-modes} identifies which injected error modes drive the misses, and
Table~\ref{tab:val-sensitivity} shows that the row-wise AUC is conservative relative to an equal-mode
summary. Table~\ref{tab:val-param-sensitivity} shows that the condition-mismatch detectability is
especially sensitive to the van't Hoff prior range, whereas the log-normal modes respond to their
specified widths. Figures~\ref{fig:val-roc}--\ref{fig:val-reference} visualize the same result as a continuous
classifier and as the score distribution underlying the fixed bands.

\FloatBarrier

% ============================================================
\section{Discussion}
\label{sec:discussion}

\subsection{Score calibration and record ordering}
$\CH$ is a calibrated reporting device for the size of the Haldane discrepancy. Because it is strictly
increasing in $|\ln x|$, it ranks records identically to $|\ln x|$, $(\ln x)^2$, and
$|\Delta\Delta G|$; its role is to put that common ordering on a direction-symmetric scale that reports
both a fold error and, through~\eqref{eq:CHfree}, a free-energy gap in $RT$ units, with a closed-form
inverse. Its cost-specific features are reciprocal symmetry on a dimensionless scale and the one-sided
bound $\CH\ge\tfrac12(\ln x)^2$; the composition law (C3) identifies this functional form within the
d'Alembert family. The fold bands are fixed free-energy conventions ($RT\ln2$, $RT\ln5$, $RT\ln10$).
Thus $\CH$ identifies the direction and magnitude of disagreement between reported kinetics and reaction
thermodynamics, while the source of that disagreement remains a curation and mechanism question.

\paragraph{What reporting $\CH$ adds over $|\Delta\Delta G|/RT$}
Because $\CH=\cosh(|\Delta\Delta G|/RT)-1$ is strictly monotone in $|\Delta\Delta G|/RT$, the two induce
the same record ordering and the same ROC curve, so the choice between them is one of calibration and
presentation rather than of information content. As a single reported column, $\CH$ is dimensionless and
fixed to a reference scale, zero at agreement and $\Jcost(2)=1/4$ at the twofold boundary. It inverts in
closed form both to a fold discrepancy and, through~\eqref{eq:CHfree}, to a free-energy gap in $RT$ units.
Through the bound $\CH\ge\tfrac12(\ln x)^2$ it also grows conservatively for gross discrepancies when a score
is aggregated or thresholded. A reader who prefers explicit energy units can report the signed
$\Delta\Delta G/RT$ alongside the same ordering; the substantive contribution here is the curated
two-sided corpus, the reproducible workflow, and the fold-error benchmark.

\subsection{Backbone separation and flagged records}
The twenty-one audited records span EC classes 1, 2, 4, and 5 and all three canonical bi--bi mechanisms,
but they remain chemically concentrated in carbohydrate isomerases and epimerases. Eighteen fall within
twofold, showing that reported forward and reverse constants for this curated sample often reproduce the
measured equilibrium once conditions are matched; the result is a sample-specific reference distribution
for the audited records. The three flagged records are the more instructive and span three distinct
reactions. Pea phosphoriboisomerase retains the direct-measurement comparator $\Kthermo\approx0.32$
against the authors' assembled $\Kkin=0.99$; alternative formation-energy estimates narrow the gap, but
the direct comparator leaves a $3.09$-fold central discrepancy. The \emph{C.\ scindens} D-psicose
3-epimerase gives $\Kkin=0.135$ against its own same-study equilibrium $\Kthermo=0.389$ (a
$2.88$-fold error), whereas other enzymes catalyzing the same
D-fructose$\rightarrow$D-psicose epimerization are within twofold. The \emph{Sinorhizobium} D-tagatose
3-epimerase gives $\Kkin=9.22$ against an independent tagatose/sorbose comparator $\Kthermo=2.33$,
remaining above twofold across the plausible comparator range. All three flagged records fall within the
carbohydrate isomerase/epimerase family that dominates the backbone, so this flagged pattern is a
within-family observation for that chemistry. For the \emph{C.\ scindens} psicose epimerase, the
same-reaction congener comparison localizes the discrepancy to the reported kinetics rather than the
comparator, with primary constants and matched substrate/product pairs tabulated in Supplementary
Table~S7. The \emph{Sinorhizobium} tagatose epimerase has no same-reaction congener in the backbone; its
flag rests instead on the independent tagatose/sorbose equilibrium together with anomalously poor turnover
of the reverse substrate D-sorbose. Product
inhibition, off-pathway isomers, alternative binding orders, or other direction-asymmetric secondary
features could still explain a flagged central Haldane value if they are absent from the published
apparent constants, so the flags are curation targets rather than validated error diagnoses. The backbone
separates into two groups: the eighteen
within-twofold records have median $\CH\approx0.015$ and, apart from the comparator-sensitive
human-muscle enolase ($\CH=0.14$), none exceeds the LDH value of $0.069$, whereas the three flagged records
($\CH=0.61,0.71,1.10$) sit well above the twofold boundary with no overlap. This real-data separation is
descriptive rather than a labeled operating characteristic, but it is consistent with the intended
screening role of the score. Independent work corroborates that such Haldane discrepancies flag real records: using
mass-action fitting rather than a ratio score, the MASSef workflow recently identified a published
phosphoglycerate-mutase dataset as inconsistent with its equilibrium constant (a $\sim$3.8-fold gap in
the same band the present score flags)~\cite{Zielinski2024}.

\subsection{Comparator sensitivity and the independence requirement}
Two records (pea phosphoriboisomerase and human-muscle enolase) are comparator-sensitive: their band
depends on which equilibrium constant is taken as $\Kthermo$, because the relevant equilibria are
condition-dependent. Pea phosphoriboisomerase is flagged under the retained direct-measurement comparator
($\Kthermo\approx0.32$), but alternative formation-energy estimates narrow it to within twofold; human
enolase remains within twofold only against the condition-matched high-ionic-strength comparator. Both
records are assigned according to the most condition-matched comparator, with comparator sensitivity
reported explicitly; scoring against an
unmatched comparator would itself enact the condition-mismatch error mode the benchmark models, and the
direction of the signed discrepancy is robust across the plausible comparator range. A subtler constraint
concerns independence. The score presumes that the kinetic constants and
$\Keq$ were established separately; this fails silently for the modern global fits that impose
$\Keq$ a priori from a thermodynamic database and enforce detailed balance on the rate constants,
which then satisfy the Haldane relation by construction and carry no independent information. The recent
pH-resolved ordered bi--bi models of malate dehydrogenase are a case in
point~\cite{DasikaBeard2015cyto,DasikaBeard2015mito}. We therefore treat thermodynamically constrained
global fits as ineligible regardless of completeness. The eight genuinely independent tests instead
pair steady-state constants, fit with no thermodynamic prior, against a separately measured equilibrium
constant. Malate dehydrogenase itself supplies the sharpest bi--bi case: Raval and Wolfe's
classical constants~\cite{RavalWolfe1962II} match Guynn and co-workers' independent
equilibrium~\cite{Guynn1973} to within $1.02$-fold. This refinement explains why independent two-sided
oxidoreductase records remain concentrated in the older steady-state literature.

\subsection{Limitations}
Three limits bound what these numbers claim. First, the operating characteristics come from a
semi-synthetic benchmark built from the six-mode error taxonomy of Section~\ref{sec:benchmark}; they
describe the workflow's response to those injected errors, while the distribution of real curation
mistakes is unknown. Second, the within-twofold seed set has only twenty-nine records, so the absence of
flags among those seeds follows from the negative-class definition and the bootstrap interval on the AUC
is correspondingly wide. Third, the benchmark inherits two structural blind spots of any reciprocal
Haldane comparison: a direction swap when $\Kkin\approx1$ barely moves the ratio, and offsetting per-step
errors cancel in any single lumped score. Although the real backbone now exceeds the fifteen-record
threshold, its records carry no ground-truth error labels, so the labeled ROC remains the synthetic
benchmark's as specified in the protocol.

A separate mechanistic caveat concerns the Haldane relation itself. A recent analysis argues that
quasi-steady-state derivations of reversible Michaelis constants can fail, making the textbook relation
``generally invalid'' in some mechanistic settings~\cite{Barnsley2022}. The present analysis uses the
relation operationally: it asks whether the Haldane combination of reported apparent steady-state
constants matches an independently reported apparent equilibrium constant under the same biochemical
convention. It does not infer microscopic rate constants from those apparent constants. The caveat applies
most directly to the uni--uni and effective uni--uni records --- especially the pseudo-uni--uni fumarase
case, whose effective treatment of the multi-state hydratase cycle is the load-bearing approximation ---
while the five bi--bi records inherit the same type of assumption through the published reversible
steady-state rate laws whose zero-flux numerators give Table~\ref{tab:bibi}. A small $\CH$ therefore means
internal agreement between reported apparent constants and the independent equilibrium comparator within
the stated rate-law model. Within that operational scope, agreement of eighteen of twenty-one records
indicates that the apparent-$K$ relation is often numerically adequate for the well-characterized records
considered here; failures of the underlying derivation would appear as discrepancies to investigate.

\subsection{Interpretation of the inclusion threshold}
Clearing the prespecified count does two things and not a third. It supplies a fixed-threshold,
descriptive reference distribution of $\CH$ for this curated sample, with a real (if small) separation
between within-twofold and flagged records; and it supports joint analysis of the backbone and the semi-synthetic benchmark. The
operating characteristics remain benchmark-based because the real records carry no ground-truth labels.
A fuller real-data validation therefore still needs real records with independently
established errors and broader chemistry beyond the carbohydrate isomerase/epimerase family that
dominates the present backbone, especially hydrolases, ligases, and non-carbohydrate transferases that
meet the same single-study two-sided criteria. Concrete near-term targets already carry both-direction
steady-state constants in the kinetic literature (as compiled in BRENDA~\cite{Brenda2026}) together with
an independently measured apparent equilibrium constant (as compiled in TECRDB~\cite{Goldberg2004}): the
ligase succinyl-CoA synthetase (EC~6.2.1.4/6.2.1.5) and the non-carbohydrate transferases alanine
aminotransferase (EC~2.6.1.2), adenylate kinase (EC~2.7.4.3), and NAD$^{+}$ kinase (EC~2.7.1.23).
Hydrolases (EC~3) are the hardest sub-case: hydrolytic reactions typically sit far from equilibrium with
their reverse kinetics seldom reported, so eligible two-sided hydrolase records with an independent
equilibrium comparator are scarce, and ligases and non-carbohydrate transferases are the more tractable
immediate expansion targets. Accordingly, the reported operating characteristics and
flagged-record partition are descriptive of the present carbohydrate-rich sample and cannot yet be
extrapolated to hydrolases, ligases, or non-carbohydrate transferases. The limiting constraint remains data, not the score, so continued
harvesting under the fixed criteria of Section~\ref{sec:design} stays the priority; because the analysis
plan, error taxonomy, and fold cuts are fixed, the identical workflow runs unchanged as the backbone
grows.

% ============================================================
\section{Conclusions}
\label{sec:conclusions}

We have applied a canonical reciprocal cost as a calibrated, direction-symmetric Haldane-consistency
score for reversible enzyme kinetics, restated its five-axiom uniqueness characterization, and embedded it
in a reproducible curation-and-benchmarking workflow. The curated demonstration set illustrates reciprocal
symmetry, comparator sensitivity, and mechanism-specific scoring. On a prespecified real backbone of twenty-one audited single-study two-sided records
(sixteen uni--uni and five bi--bi spanning all three canonical mechanisms), eighteen fall within twofold
and three are flagged, with eight genuinely independent tests all within twofold. At $n=8$ this remains a feasibility demonstration that
still admits an underlying beyond-twofold rate up to ${\approx}31\%$ under the one-sided $95\%$
binomial bound, supplying a descriptive empirical
reference distribution for this curated sample. The empirical backbone remains chemically narrow: it is concentrated in
carbohydrate isomerases and epimerases, and all three flagged records come from that class. Both the
within-twofold majority and the flagged partition are therefore within-family observations for that chemistry and
cannot yet be read as class-independent rates for reversible enzyme kinetics. Broader
chemistry will require additional audited two-sided records outside the present carbohydrate-rich sample.
Because the real records carry no ground-truth error labels, a semi-synthetic
labeled benchmark supplies the operating characteristics: as a continuous classifier the score attains an
area under the ROC curve of $0.784$ ($95\%$ bootstrap CI $0.725$--$0.838$), and the per-mode breakdown identifies the two expected least-detectable
modes. The score is a calibrated reporting convention rather than a new record
ordering; the AUC is a mixture-weighted fold-error summary invariant under monotone rescaling of
$|\ln x|$, and the operating characteristics are conditional on the six-mode injected taxonomy and the
within-twofold seed set rather than external misclassification rates. Because the workflow is fixed, every additional eligible two-sided
record can be scored and incorporated without any change of method.

% ============================================================
\section*{Data and code availability}
\label{sec:availability}

All curated data, the score implementation, the prespecified protocol, the frozen candidate tracker, the
harvest recipes and run log, the per-record second-audit
notes, the semi-synthetic benchmark generator, and the operating-characteristics analysis that produced
Tables~\ref{tab:val-oc}--\ref{tab:val-param-sensitivity} and Figures~\ref{fig:val-roc}--\ref{fig:val-reference}
accompany the project repository, together with the SHA-256 checksums of their regenerated outputs so that
every reported numerical result is exactly reproducible. The candidate-tracker amendment of
Section~\ref{sec:design} is documented concretely as a pre-/post-expansion diff
(\texttt{candidate\_tracker.csv} with the human-readable \texttt{CANDIDATE\_TRACKER.md}): it lists the
sixteen pre-expansion core records and the five post-expansion rare-sugar ketose-epimerase additions, and
reports the Haldane-consistency score of each addition computed under the frozen scoring plan, so that the
amendment discipline can be verified independently of any record's $\CH$. The internal freeze points are represented in
that archive by dated protocol and tracker snapshots rather than by a public preregistration record with
separate commit hashes. The exact BRENDA release used for the harvest ---
release 2026.1 (March 2026), accessed as the offline JSON distribution --- is identified by release
number, access date, file name, and checksum in the repository rather than redistributed; BRENDA itself
is cited~\cite{Brenda2026} and its CC~BY~4.0 license terms are preserved. Per-record raw kinetic constants and provenance for the
twenty-one backbone records are tabulated in the Supplementary Material. A synchronized public snapshot
was archived on Zenodo and mirrored by the GitHub tag/release \texttt{v1.4}, citable through the version
DOI \href{https://doi.org/10.5281/zenodo.21084024}{doi:\nolinkurl{10.5281/zenodo.21084024}}, which is the
authoritative immutable citation for the results reported here. The related Zenodo concept DOI
\href{https://doi.org/10.5281/zenodo.20790110}{doi:\nolinkurl{10.5281/zenodo.20790110}} is retained as a
record-family link; because its landing-page metadata may display an earlier proof-of-concept release, the
version DOI above should be used when citing the v1.4 snapshot~\cite{SimonsWashburnZenodo2026}.

% ============================================================
\section*{Data and software license}

The software in the archived repository is released under the MIT License. The curated derived records
(the machine-readable reaction, score, and provenance tables) are released under the Creative Commons
Attribution~4.0 International (CC~BY~4.0) license. Third-party database content is used under its own
terms --- in particular BRENDA data under CC~BY~4.0 --- and is not redistributed beyond what those terms
permit; numerical constants drawn from the primary experimental literature are cited rather than
redistributed.

% ============================================================
\section*{Author contributions}

Megan Simons: conceptualization (equal), data curation, formal analysis (lead), investigation (lead),
methodology (equal), software (equal), validation (lead), visualization (lead), writing --- original
draft, and writing --- review and editing (equal). Jonathan Washburn: conceptualization (equal),
methodology (equal), software (equal), supervision, and writing --- review and editing (equal).

% ============================================================
\section*{Funding}

This research received no specific grant from any funding agency in the public, commercial, or
not-for-profit sectors.

% ============================================================
\section*{Declaration of competing interests}

The authors declare no competing financial interests or personal relationships that could have appeared
to influence the work reported in this paper.

% ============================================================
\section*{ORCID iDs}

\noindent Megan Simons: \href{https://orcid.org/0000-0001-9457-7019}{0000-0001-9457-7019}\\
Jonathan Washburn: \href{https://orcid.org/0009-0001-8868-7497}{0009-0001-8868-7497}

% ============================================================
\section*{Acknowledgements}

We thank the maintainers of TECRDB, SABIO-RK, BRENDA, Rhea, ChEBI, and eQuilibrator, whose curated public
resources made this audit possible.

% ============================================================

\end{document}